\documentclass[12pt]{article}
\usepackage{setspace}
\usepackage{tikz}
\usepackage{float}
\usepackage{svg}
\usepackage{ragged2e}
\usepackage{pdflscape}
\usepackage{comment}
\usepackage{amsmath}
\usepackage{stackengine}
\usepackage{amsfonts}
\usepackage{amssymb}
\usepackage[citestyle=authoryear, bibstyle=authoryear, natbib=true, backend=biber, maxnames=2, uniquename=false]{biblatex}
\usepackage{graphicx}
\usepackage{subcaption}
\usepackage{authblk}
\usepackage[colorlinks=true,urlcolor=black,citecolor=black,linkcolor=black,bookmarks=true]{hyperref}
\usepackage{booktabs}
\usepackage[flushleft]{threeparttable}
\usepackage{multirow}
\usepackage{vmargin}
\setmargins{2.0cm}       
{1.5cm}                  
{16.5cm}                 
{23.42cm}                
{10pt}                   
{0.5cm}                  
{0pt}                    
{2cm}                    

\usepackage{listings}
\usepackage{color} 
\definecolor{mygreen}{RGB}{28,172,0} 
\definecolor{mylilas}{RGB}{170,55,241}

\bibliography{references.bib}

\usepackage{fancyhdr}

\begin{document}
\lstset{language=Matlab,%
    breaklines=true,%
    morekeywords={matlab2tikz},
    keywordstyle=\color{blue},%
    morekeywords=[2]{1}, keywordstyle=[2]{\color{black}},
    identifierstyle=\color{black},%
    stringstyle=\color{mylilas},
    commentstyle=\color{mygreen},%
    showstringspaces=false,
    numbers=left,%
    numberstyle={\tiny \color{black}},
    numbersep=9pt, 
    emph=[1]{for,end,break},emphstyle=[1]\color{red}, 
}

\begin{singlespace}

\setstretch{1.3}
\title{ \textbf{Chapter 1: Labor Market Reforms, Flexibility, and Employment Transitions Across Formal and Informal Sectors}}

\author[]{Selidji Tossou}

\maketitle

\begin{abstract}

In this paper, I investigate the 2017 labor market reform in Benin, which reduced firing costs and allowed firms to renew short-term contracts indefinitely. Using micro-data from the Harmonized Surveys on Households Living Standards and a two way fixed effect approach with nearby countries as the control group, I assess the reform’s impact on employment, worker tenure, contract types, and wages. My empirical results reveal a 2.6 percentage point (24.5\%) increase in formal sector employment and a 2.8 percentage point (3.2\%) reduction in informal employment. Formal sector tenure decreased by 0.23 months for short-term contract workers, reflecting higher turnover, while long-term contract tenure increased by 0.15 months. The likelihood of securing a permanent contract rose by 23.2 percentage points (41.6\%) in the formal sector, indicating that firms used long-term contracts to retain high-productivity workers. Wages in the formal sector increased by 33.6 USD per month on average, with workers on short-term contracts experiencing a wage increase of 19.6 USD and those on long-term contracts seeing an increase of 23.4 USD. I complement these findings with a theoretical job search model, which explains the mechanisms through which lowered firing costs affected firm hiring decisions, market tightness, and the sorting of workers across sectors. This study provides robust evidence of labor market reallocation and highlights the complex trade-offs between flexibility, employment stability, and wages in a developing country context.

\end{abstract}

\end{singlespace}

\clearpage

\section{Introduction\label{introduction}}

Labor market regulations are critical in shaping employment outcomes, especially in developing economies with high levels of informality. Benin, where 8 in 10 individuals of the workforce operates within the informal sector, stands as a prominent example of such an economy. The informal sector, accounting for an estimated 60-70\% of GDP, is a major driver of economic activity in the country \citep{worldbank2023, afdb2022, ilo2017}. The formal sector, on the other hand, remains relatively small but heavily regulated, with stringent labor laws and high firing costs limiting the flexibility of firms. Amid persistent poverty, with 36.2\% of the population living below the poverty line, and significant underemployment rates (72\% of the workforce), understanding how labor market reforms influence both formal and informal employment is essential to fostering sustainable growth and job security \citep{instad2022, worldbank2023}.

In 2017, Benin enacted a comprehensive labor market reform aimed at increasing labor market flexibility by reducing firing costs—particularly for long-term contracts, and removing the legal limit on the use of short-term contracts (STCs) in the formal sector. Previously, firms were only allowed to renew STCs for a maximum cumulative duration of 48 months, after which they were required to offer a long-term contract or terminate the worker. The reform eliminated this cap, allowing firms to use STCs indefinitely. Short-term contracts refer to employment agreements that do not guarantee long-term job security, allowing employers to hire workers for limited, often renewable, periods without the commitment of permanent employment. In contrast, long-term contracts provide stable, ongoing employment with greater job security, typically including benefits and severance requirements that make dismissal more complex and costly for employers. Prior to the reform, firms in the formal sector were constrained by high severance payments and complex dismissal procedures, which limited their ability to adjust their workforce in response to changing economic conditions. The reform's goal was to stimulate job creation by providing firms with more flexibility in hiring and firing decisions \citep{tossou2025essays}. However, the consequences of such reforms are multifaceted. While they may encourage firms to expand formal sector employment, there is also the potential for increased turnover, reduced job tenure, and greater reliance on short-term contracts—both because of the newly granted legal flexibility and because they help firms manage dismissal costs and uncertainty in worker productivity. Additionally, such reforms might have spillover effects on the informal sector and worker mobility.

This paper analyzes the impact of the 2017 labor market reform on employment dynamics, contract types, job tenure, and wages using a two way fixed effect approach. The analysis leverages data from the Harmonized Surveys on Household Living Standards to compare labor market outcomes in Benin before and after the reform with those in other West African countries that did not undergo similar reforms. Prior to examining the reform’s effects, I ensured balance in baseline characteristics across treatment and control groups and conducted tests for parallel trends in key employment outcomes. These checks confirmed stable pre-reform trends, supporting the validity of the two way fixed effect approach. By focusing on the reallocation of labor between the formal and informal sectors, the paper provides new insights into how labor market flexibility influences both sectors and highlights the complex trade-offs between job creation and job security.

The results indicate that the reform led to significant changes in employment patterns. Formal sector employment increased by 18.57\% (a 2.6 percentage point increase), while informal sector employment decreased by 3.36\% (a 2.8 percentage point reduction). These results suggest that the reduction in firing costs enabled firms to expand formal employment, drawing workers away from the informal sector. However, overall employment remained stable, with the probability of working at all decreasing marginally by 0.10\% (-0.147 percentage points, non significant), indicating that the reform primarily facilitated a reallocation of workers between sectors rather than a significant net increase in employment.

A key outcome of the reform was the shift in contract types within the formal sector. The probability of having a permanent (long-term) contract in the formal sector increased by 23.2 percentage points (41.6\%), and the overall probability of obtaining a permanent contract rose by 4.6 percentage points (88.4\%). This shift was particularly pronounced among specific demographic groups, with long-term contracts increasing by 49.7\% for female workers, 59.5\% for rural workers, and 57.6\% for unmarried workers. These results highlight the firm’s response to increased labor market flexibility by offering permanent contracts to high-productivity workers to reduce turnover and retain skilled employees. 

Despite these positive developments, the reform also led to a reduction in job tenure, particularly for workers on short-term contracts (STCs). Empirical results show that tenure for short-term contract workers in the formal sector decreased by 0.23 months, reflecting higher turnover among low-productivity workers. In contrast, tenure for workers with long-term contracts increased slightly by 0.15 months, as firms sought to retain valuable employees through more stable employment arrangements. This reduction in job tenure can be attributed to two effects: a composition effect, where the influx of new workers entering the formal sector altered the overall worker composition, and a direct impact of the reduced firing costs, which facilitated the dismissal of low-productivity workers previously retained under long-term contracts. On aggregate, job tenure across all workers decreased by 0.16 months, signaling a broader trade-off between labor market flexibility and job stability.

Wage dynamics also shifted in response to the reform. The average wage in the formal sector increased by 33.6 USD per month, with long-term contract workers benefiting from a wage increase of 23.4 USD per month. Short-term contract workers experienced a more modest wage increase of 19.6 USD per month. These wage gains can be attributed to firms’ ability to negotiate higher wages for high-productivity workers while managing labor costs through flexible employment arrangements. The theoretical model developed in this paper provides an explanation for these wage dynamics, showing that lower firing costs allow firms to offer more competitive wages to retain skilled workers in a tight labor market.

Theoretical insights from an extended Diamond-Mortensen-Pissarides (DMP) model help explain the mechanisms driving these empirical results. The model incorporates endogenous job separation rates, worker-firm match quality, and sectoral differences in labor market regulations, offering a nuanced view of how labor market reforms influence employment outcomes. By lowering firing costs, the reform increased the separation threshold, \( \gamma_F^{\text{min}} \), leading to higher formal sector vacancies and more frequent dismissals of low-productivity workers. At the same time, firms in the formal sector responded to increased competition for skilled workers by offering more permanent contracts and higher wages, as predicted by the Nash bargaining framework in the model. These theoretical insights align closely with the observed empirical results, illustrating how labor market flexibility reshapes the allocation of workers across sectors and alters firm hiring strategies.

In summary, the 2017 labor market reform in Benin successfully increased formal sector employment, boosted the prevalence of long-term contracts, and raised wages, particularly for high-productivity workers. However, the reform also resulted in higher turnover rates and shorter job tenures for workers on short-term contracts, highlighting the challenges of balancing labor market flexibility with job security. Policymakers may consider complementary measures, such as retraining programs and enhanced unemployment benefits, to mitigate the negative effects of short-term contracts on vulnerable workers. Future reforms can aim to strike a balance between increasing labor market flexibility and ensuring stable employment opportunities, particularly in economies with high levels of informality. This paper offers both empirical and theoretical contributions to the literature on labor market reforms, providing valuable insights for policymakers and researchers interested in promoting inclusive labor market growth. The subsequent sections are organized as follows: Section \ref{literature} reviews the literature on the effects of labor market flexibility reforms; Section \ref{background} outlines the 2017 reform in Benin; Section \ref{method} presents the data and methodology; Section \ref{results} discusses the empirical results; Section \ref{model} introduces the theoretical model and its implications; and finally, Section \ref{conclusion} concludes with policy recommendations.


\section{Review of the Literature\label{literature}}

The theory of job matching and worker turnover, first formalized by \textcite{jovanovic1979job}, has long served as the foundational framework for analyzing labor market dynamics. This seminal theory inspired a vast literature in applied microeconomics, particularly in studying how employment protection legislation influences turnover and employment stability. Empirical and theoretical work has explored the repercussions of such protections on labor market performance, especially in the European context, where employment legislation is more rigid \parencite{hijzen2013perverse, noelke2017job, feldmann2009effects, kahn2012labor}. These studies typically show how stringent job security regulations reduce turnover and employment in European countries. However, labor market flexibility reforms, which reduce hiring and firing costs, have been shown to have mixed effects on labor markets across various contexts \citep{bertola1992labor, aguirregabiria2014labor}.

The labor market rigidity observed in many European countries contrasts sharply with North American experiences, where more flexible labor markets contribute to lower unemployment rates \citep{aguirregabiria2014labor}. Several studies point to severance payments as a critical factor explaining these differences, as severance pay reduces dismissals during economic downturns but also deters hiring during expansions \citep{aguirregabiria2014labor}. The net effect of firing costs on employment depends on a variety of factors, including labor market frictions and the persistence of demand shocks. In this context, \citet{bertola1992labor} theorized that firing costs might increase employment under certain conditions, even if hiring costs reduce it, depending on the firm’s discount rate and attrition. Evidence from Colombia, for example, demonstrates that reducing dismissal costs significantly increases transitions into and out of unemployment, particularly for workers in formal enterprises compared to their informal counterparts \parencite{kugler1999impact}.

However, the empirical evidence from Europe and the United States on severance payments and job protection laws is often ambiguous and mixed, with conflicting conclusions about their overall impact on employment outcomes. This body of research largely ignores developing countries, where labor market dynamics are often dominated by informal employment, and data availability is limited. To my knowledge, no prior study has examined the effects of labor market reforms in Benin, making this paper one among the first to assess the implications of labor market flexibility reforms in the African context. The existing literature on European and Latin American countries does not translate easily to developing countries like Benin, where the informal sector plays a far more significant role in shaping labor market outcomes. In Benin, 8 in 10 employments were informal between 2011 and 2017, and the informal sector contributes 60-70\% of GDP \citep{ilo2017, worldbank2023}. These distinct characteristics necessitate an independent examination of how labor market reforms affect employment in developing economies.

Recent research has expanded on labor market flexibility reforms in other African countries. For instance, a study by the IMF examines South Africa’s labor market reform options, highlighting that reducing dismissal costs and relaxing hiring restrictions could help address the country’s high unemployment rates \parencite{IMF2021SouthAfrica}. Additionally, the World Bank has analyzed how trade liberalization impacted South African labor markets, observing that tariff reductions led to shifts from formal to informal employment as workers adapted to changing market conditions \parencite{worldbank2023southafrica}. Finally, \textcite{kar2016reforms} discusses how economic reforms, such as reducing hiring and firing restrictions, often increase formal employment flexibility while expanding informal employment as workers shift to less regulated sectors. These studies provide insights into how labor market reforms can variably impact employment dynamics across African economies with significant informal sectors.

This paper bridges the gap in the literature by analyzing the labor market effects of flexibility reforms in a country where informality dominates the labor market. The findings contribute to broader debates on labor market regulation, employment protection, and flexibility, offering insights that are relevant not only for Benin but for other developing countries facing similar labor market challenges.

\section{Background\label{background}}

In Benin, changes to labor market regulations were implemented in 2017, fundamentally altering the structure of employment contracts and dismissal processes in the formal sector. The reform marked a significant shift by reducing severance pay for workers under long-term contracts, extending the use of short-term contracts, lowering compensation for unfair dismissal, and broadening the legal grounds for both individual and economic dismissals. These changes sharply reduced the firing costs for firms in the formal sector, a substantial departure from the previous regulatory framework established in 1998. By redefining core elements of the employer-employee relationship, this reform represented a pivotal move toward increased labor market flexibility and aimed to reshape the employment landscape for firms and workers alike.

The formal sector in Benin is subject to taxation and regulation, with regular wages and explicit contracts between employees and employers. Firms operating in this sector benefit from legal protections, access to formal credit, and participation in public contracts. These advantages, coupled with the potential for long-term productivity and growth, provide strong incentives for firms to formalize their operations. Additionally, formal contracts allow firms to attract and retain skilled workers, who may prefer the stability and benefits of formal employment. The government intended these reforms to encourage formalization by lowering the costs associated with formal employment, thereby broadening the tax base and increasing tax revenues. Furthermore, by reducing labor contract rigidity, the government aimed to foster job security and build a stable and productive workforce — at least relative to informal employment, where workers typically lack enforceable contracts, severance protections, and access to social security.

Conversely, the informal sector, which comprises mainly small firms or self-employment, is not subject to institutional regulation. With informality rates particularly high across Sub-Saharan Africa, Benin is no exception. According to estimates from the National Statistics Agency in 2009, the informal sector accounted for as much as 70 percent of GDP. This prevalence of informality poses challenges for governments, which struggle with tax collection, limiting their ability to fund public services. Informal enterprises also lack access to formal financial resources, public contracts, and governmental support programs, making them vulnerable to corruption or pressure from authorities. Formal businesses, in turn, face competition from the informal sector and may bear a disproportionate tax burden \citep{worldbank2015finding}.

Benin’s recent reforms build upon broader economic liberalization efforts that began in 1990, when the country embraced economic liberalism and engaged in financial reforms with support from the World Bank and the International Monetary Fund. Significant changes to the country’s management and economic policies helped restore macroeconomic stability and strengthen growth, gradually improving the macroeconomic framework and consolidating public finances.

\begin{figure}[H]
    \centering
    \caption{Old vs. New Labor Regulation in Benin: Key Changes}
    \label{policy}    
    \includegraphics[width=\textwidth]{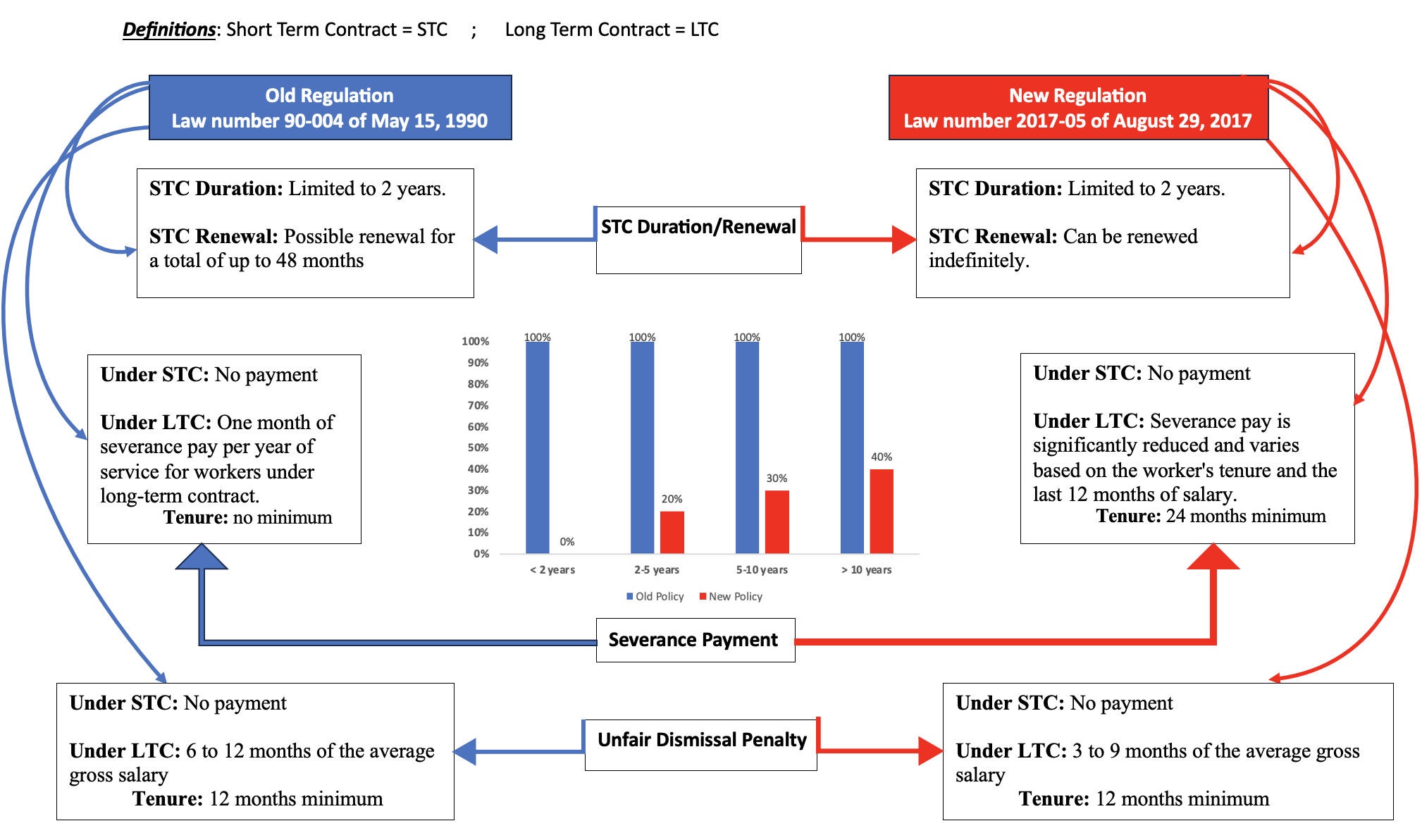}
    \vspace{0.5cm} 
    \begin{minipage}{\textwidth}
        \footnotesize{
        \justifying
        \footnotetext{This figure compares the key differences between Benin's old labor regulation (Law 90-004 of May 15, 1990) and the new labor regulation (Law 2017-05 of August 29, 2017). It highlights changes to the duration and renewal of short-term contracts (STC), severance pay, and unfair dismissal penalties. Under the new law, the renewal of STCs became indefinite, and severance pay for long-term contracts (LTC) was significantly reduced, particularly for workers with fewer years of service. This led to decreased costs for employers and increased labor market flexibility. Source: By Author.}
        }
    \end{minipage}
\end{figure}

Law 98-004 of January 27, 1998, governed workforce declarations, hiring, and contract termination until the 2017 reforms. The updated rules, implemented under law 2017-05 of August 29, 2017, introduced new conditions for hiring, placement, and termination of employment contracts. Key changes included the reduction of firing costs by lowering compensation for unfair dismissal, allowing indefinite renewals of short-term contracts (STCs), and decreasing severance pay upon dismissal. Each of these changes affects workers differently based on the type of contract held, as summarized in Figure \ref{policy}.

The reform’s extension of STCs was a significant shift. Previously, under Law 90-004 of May 15, 1990, STCs were limited to a maximum duration of two years, with renewal allowed up to a total of 48 months. Employers then had to either convert these contracts to long-term employment or let go of the worker. However, under the 2017 reform, STCs can now be renewed indefinitely, allowing employers to maintain short-term arrangements without transitioning workers to long-term contracts. This flexibility reduces firing costs, as termination of an STC does not require severance pay. Employers are required to give a two-month notice to avoid penalties. This change is expected to increase turnover in the formal sector.

The reform also reduced severance pay for dismissed workers. Previously, workers under long-term contracts (LTCs) were entitled to one month of severance per year of service with no minimum tenure requirement. Following the reform, severance pay now varies by tenure and salary, with a minimum of 24 months of service required, reducing the overall amount paid out in severance. 

Finally, the penalty for unfair dismissal was also reduced. Under the old law, compensation ranged from 6 to 12 months of average gross salary from the previous year. This amount was reduced to 3 to 9 months, and STC workers are not entitled to any compensation. Additionally, workers without at least one year of service are not eligible for compensation for unfair dismissal.

The changes in Benin’s labor market align with similar policies in other regions. For example, Spain implemented reforms that expanded the use of fixed-term employment contracts with reduced severance. Evidence from Spain suggests that these reforms increased turnover, decreased long-term employment, reduced on-the-job training, and widened wage disparities.\footnote{In Spain, labor reforms allowing for more flexible temporary contracts with lower severance pay led to a marked increase in temporary work. This expansion of temporary contracts decreased long-term unemployment but also contributed to reduced labor productivity, lower levels of on-the-job training, and increased wage inequality \citep{dolado2002drawing, toharia1993effects}. This evidence suggests potential parallels for Benin, where reforms aimed at increasing flexibility may have similar effects on turnover and productivity.}

\section{Identification strategy\label{method}}
\subsection{Two-Way Fixed Effect Method}

I exploit the temporal change in firing costs and the cross-country variation in regulatory coverage across West African countries using a two-way fixed effect model. In Benin, all workers in the formal sector are covered by the policy, while workers in the informal sector are not directly covered. However, I expect equilibrium effects to influence informal sector workers, making them unsuitable as a control group. Instead, I use a set of comparable African countries as the control group to estimate the impact of reduced firing costs on exit rates into and out of unemployment.

Let $i$ represent a worker, with two groups indexed by treatment status. Workers outside of Benin form the control group, while those within Benin constitute the treatment group. Treatment refers to being employed in Benin after the 2017 regulation change. The treatment variable $T_{c(i)t(i)}$ takes the value $1$ for a worker in Benin after the labor regulation change in 2017. Each worker is indexed as $i=1,...,N$, with $c(i)$ indicating the worker's country, $t(i)$ the observation time period (no panel data), and:
\[
T_{c(i)t(i)}= 
 \begin{cases} 
 1, & \mbox{if } c(i)= \mbox{Benin and } t(i)\geq 2017 \\ 
 0, & \mbox{otherwise} 
 \end{cases} 
\]

My main specification is a two-way fixed effect model to evaluate the treatment’s effect on an outcome variable $Y_i$, expressed as:
\begin{equation}
Y_i = \beta_0 + \beta_1 T_{c(i)t(i)} + \gamma_{c(i)} + \rho_{t(i)} + X_i^{'}\alpha + \mu_i
\end{equation}
where $\beta_0$, $\alpha$, $\gamma$, and $\rho$ are unknown parameters, and $\mu_i$ is an unobserved random error term, capturing determinants of $Y_i$ omitted from the model. The coefficient $\beta_1$ is of primary interest, as it reveals the causal effect of the treatment.

\subsection{Event Study Specification}

To examine the dynamics of the treatment effect over time, I employ an event study approach, which allows me to track the impact of the reform before and after implementation. This approach provides a visual assessment of pre-trends and helps validate the parallel trends assumption. The event study specification is as follows:
\begin{equation}
Y_{it} = \alpha + \sum_{k \neq -1} \delta_k \mathbb{1}(t = k) \times \text{Benin}_i + \gamma_{c(i)} + \rho_{t(i)} + X_i^{'}\theta + \epsilon_{it}
\end{equation}
where $\delta_k$ represents the event-time coefficients for each period $k$ relative to the reform in 2017 (with $k=-1$ as the reference period). The interaction term $\mathbb{1}(t = k) \times \text{Benin}_i$ identifies the effect of the reform in each period relative to 2017 for workers in Benin. The coefficients $\delta_k$ trace out the dynamic effects of the policy on $Y_{it}$ over time.

\subsection{Controls and Estimation Details}

I weight all regressions using household weights to account for sampling design and population representation. The model controls for urban residence, age, gender, education, household size, marital status, industry type, and poverty level, as these characteristics could affect labor market outcomes. I include fixed effects for religion and commune to account for region-specific and cultural factors that may impact employment trends. I also cluster robust standard errors at the household level to account for within-household correlation and ensure accurate inference.

\subsection{Data}

The data used to measure the effects of the new labor regulation in Benin on workers’ outcomes in the labor market are sourced from the Harmonized Surveys on Households Living Standards. The Living Standards Measurement Study (LSMS) is a key household survey initiative by the World Bank, designed to enhance household survey systems in client nations and improve the quality of microdata for more informed development policy decisions. Conducted in 2016 and 2019, these surveys are nationally representative and provide comprehensive data on poverty, employment, and other socio-economic variables across all municipalities and areas of residence (urban/rural) in Benin and other West African countries. Although the LSMS data are cross-sectional, each wave was collected progressively over several months, beginning simultaneously across all departments. I use the exact month of interview to create a monthly event-time variable.\footnote{In figures and event study analyses, month –6 corresponds to January 2016, –5 to February 2016, and so on. Month 0 represents January 2019, and subsequent months follow sequentially.} This monthly construction allows for visualization of unadjusted trends and estimation of event-time coefficients, even though individuals are not tracked over time.

I restrict the sample to individuals of working age (15–64), excluding full-time students, retirees, and others who are not part of the labor force. The final sample includes only those who are either employed or unemployed but actively seeking work at the time of the survey. The data are repeated cross-sections, and the analysis compares labor market outcomes across groups and years rather than tracking individuals over time.

The LSMS surveys offer several advantages for this study. They allow for the differentiation between formal and informal sector workers, a crucial distinction in understanding the effects of labor market reforms. In this analysis, formal sector employees are identified as those whose employers contribute to social security, while informal sector workers are those whose employers do not. This distinction closely aligns with the legal framework in Benin, where formal firms are obligated to register and declare their workforce to the social security system. As a result, the likelihood of informal workers being employed by formal firms is minimal, reinforcing the validity of this classification as a proxy for compliance with labor regulations. Additionally, the surveys provide detailed information on worker demographics—including gender, age, marital status, ethnicity, religion, education, household size, and employment industry—enabling robust controls for individual characteristics that may influence labor market outcomes such as turnover.

Employment characteristics are reported with respect to each respondent’s primary job, defined in the survey as the job that either (i) offers a formal contract or, in the absence of a contract, (ii) generates the highest income. Employment status is measured as a binary variable equal to 1 if the individual was employed at the time of the survey. Sector classification is based on the presence or absence of employer contributions to social security. Contract type is defined based on the respondent’s reported duration of the employment agreement, distinguishing between short-term and long-term contracts. Tenure refers to the number of months the respondent has spent in their primary job during the past 12 months, capped at 12 months. For individuals not currently employed, non-employment spell duration is measured in years and captures the time since the last job ended. Monthly earnings are reported in local currency and converted to USD.

\begin{figure}[H]
    \centering
    \caption{Treatment and Control Groups}
    \label{africa}
    \includegraphics[width=0.75\textwidth]{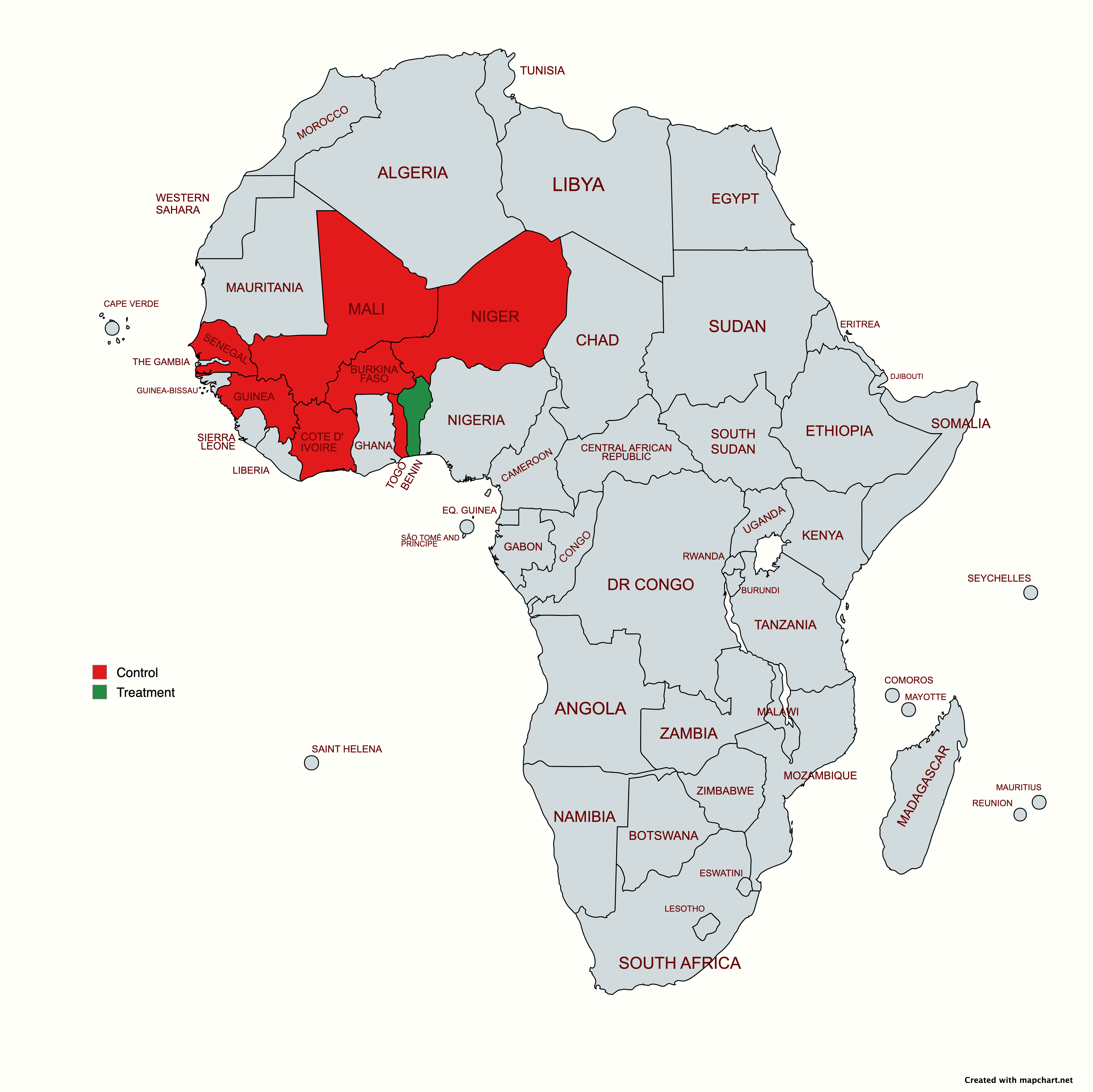}
    \vspace{0.5cm} 
    \begin{minipage}{\textwidth}
        \footnotesize{
        \justifying
        \footnotetext{This map highlights the countries used as treatment and control groups in the study. Benin, represented in green, is the treatment group where the 2017 labor market reform was implemented. The control group, marked in red, consists of West African countries that were not directly affected by the reform. These countries provide a comparative framework to assess the causal impact of the labor market reform on worker dynamics in Benin. Source: By Author.}
        }
    \end{minipage}
\end{figure}

The choice of comparison group for Benin is critical for the validity of the two-way fixed effect analysis. The comparison group consists of workers from other West African countries that are part of the West Africa Economic and Monetary Union (WAEMU) Household Survey Harmonization Project, a collaborative effort between the World Bank and the WAEMU Commission. These countries include Burkina Faso, Côte d’Ivoire, Guinea Bissau, Mali, Niger, Senegal, and Togo (Figure \ref{africa}). This selection is justified because these countries share similar economic, cultural, and institutional contexts with Benin, making them suitable as a baseline for comparison. The similarity in economic structures, labor market dynamics, and regulatory environments across these countries allows for a more accurate estimation of the causal effects of the labor reform in Benin by controlling for region-specific factors that could influence labor market outcomes. Moreover, these are the only countries available in the harmonized LSMS dataset used for this analysis, which constrains the choice of potential comparison groups but ensures consistency in survey design and measurement.

Table \ref{descriptive_stat} presents baseline summary statistics for Benin and the comparison group in 2016. Workers in Benin were on average 34.5 years old, with 47\% male, 47\% urban, and an average household size of 6.5. Educational attainment was low, with 60\% having no schooling. Table \ref{descriptive_stat_outcomes} provides additional descriptive statistics on key labor market outcomes. Approximately 10.6\% of individuals were formally employed, and 55.8\% of formal sector workers held long-term contracts. Mean tenure was 9.5 months, and average monthly earnings in the formal sector were 118.6 USD. These baseline patterns highlight structural differences between the formal and informal sectors and motivate the empirical evaluation of the reform’s effects on sectoral outcomes. All demographic characteristics summarized in Table \ref{descriptive_stat} are included as control variables in the regression analysis, along with fixed effects for industry and commune, to account for observable heterogeneity across individuals.

\section{Findings \label{results}}


\subsection{Employment}
The 2017 labor market reform in Benin, aimed at reducing firing costs and promoting the use of short-term job contracts, had diverse implications for employment across both the formal and informal sectors. The formal sector, characterized by higher regulation, responded to the reduced firing costs by giving employers greater flexibility to adjust their workforce in response to demand fluctuations. This increased flexibility likely encouraged firms to expand their workforce, as the risks associated with hiring were lowered. However, the simultaneous rise in short-term contracts created a more precarious environment for workers, increasing uncertainty around job security and benefits. This precariousness could push some workers towards the informal sector, where jobs, though less regulated, sometimes offer greater stability in certain cases but with fewer protections and benefits.

Table \ref{working} shows that the reform led to a 2.6 percentage points increase (SE = 0.004, p $<$ 0.001) in the probability of working in the formal sector, a rise of 24.5\%, while informal sector employment decreased by 2.8 percentage points (SE = 0.004, p $<$ 0.001), a reduction of 3.2\%. These shifts suggest the reform successfully reduced barriers in formal employment, particularly by lowering hiring and firing costs, encouraging workers to transition from the informal sector.

\begin{figure}[H]
    \centering
    \caption{Unadjusted Trend: Share of Employed Workers}
    \label{mean_employment}
    \includegraphics[width=0.5\textwidth]{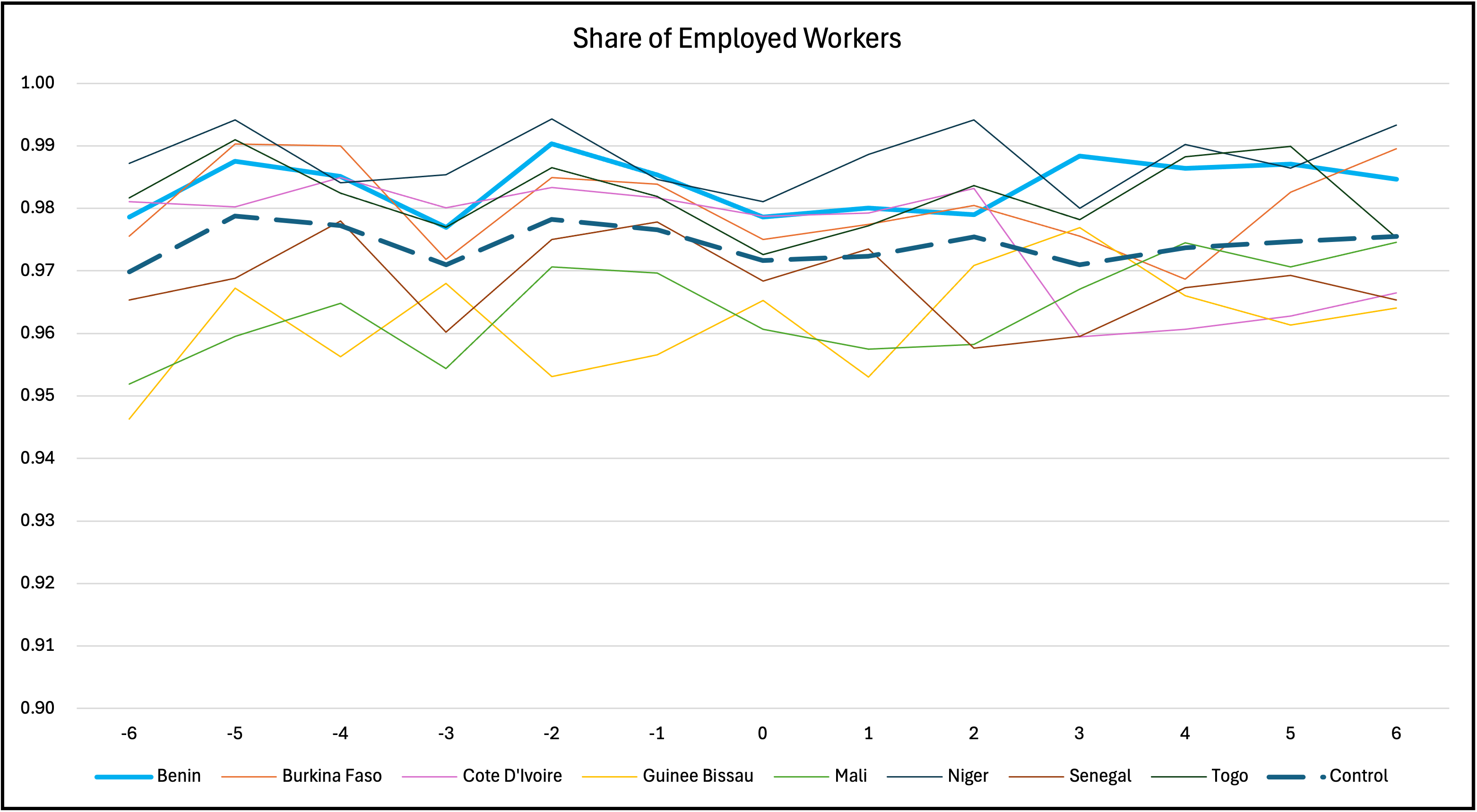}
    \vspace{0.5cm} 
    \begin{minipage}{\textwidth}
        \footnotesize{
        \justifying
        \footnotetext{This figure shows the average share of employed workers over time for Benin and several comparison countries (Burkina Faso, Cote d'Ivoire, Guinea Bissau, Mali, Niger, Senegal, and Togo). The x-axis represents pre-reform periods (–6 to –1) and post-reform periods (0 to 6), while the y-axis shows the share of employed workers. Solid lines represent individual country trends; the dotted line shows the comparison group mean. Although the LSMS data are cross-sectional, each wave was collected progressively over several months. The monthly event-time variable is based on the month of interview, with month –6 corresponding to January 2016 and month 0 to January 2019. Individuals are not tracked over time. The analysis sample includes working-age individuals (15–64) who were either employed or unemployed and actively seeking work.}
        }
    \end{minipage}
\end{figure}

\begin{figure}[H]
    \centering
    \caption{Unadjusted Trend: Share of Employed Workers in Formal Sector}
    \label{mean_employment_formal}
    \includegraphics[width=0.5\textwidth]{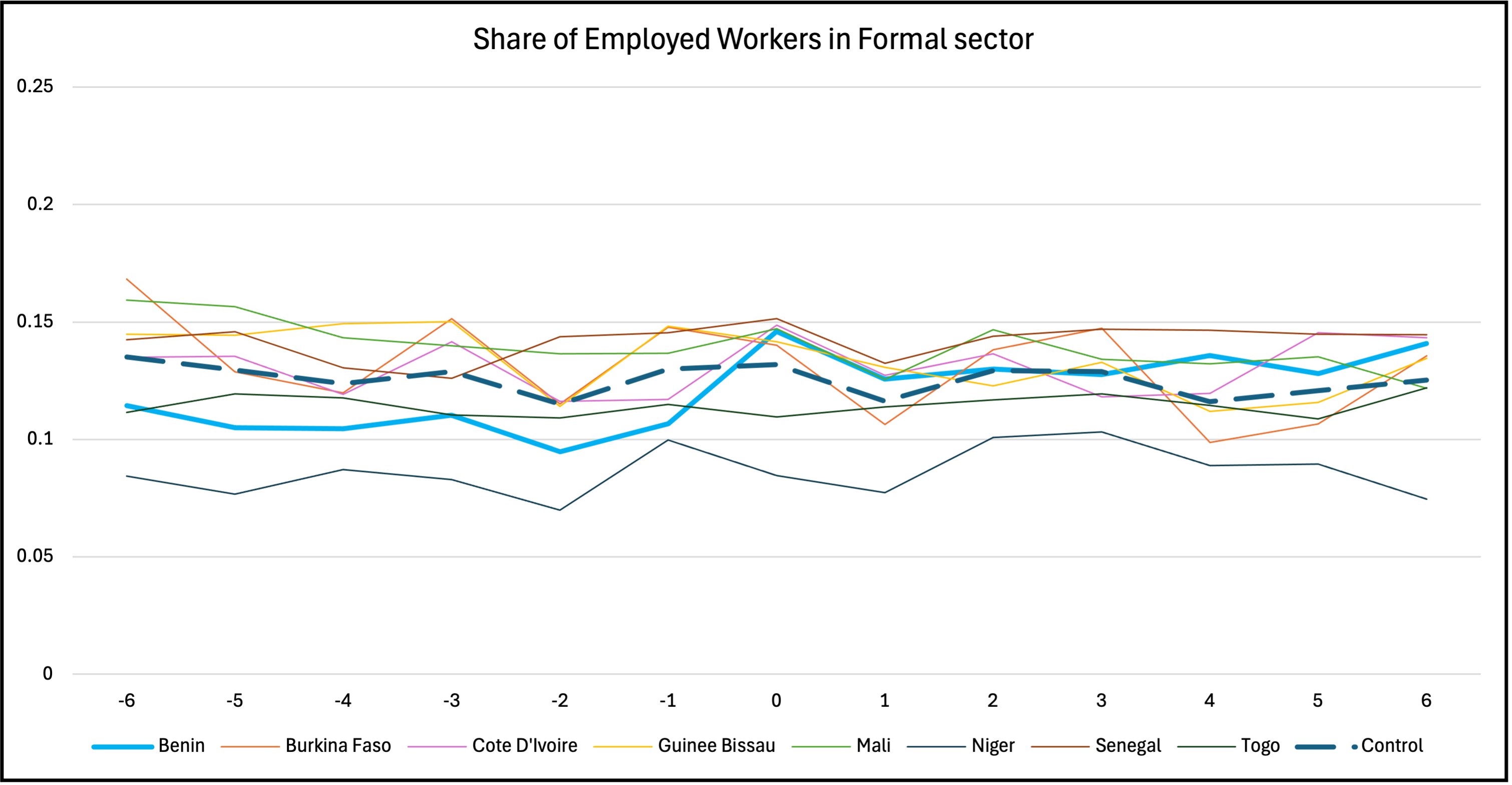}
    \vspace{0.5cm} 
    \begin{minipage}{\textwidth}
        \footnotesize{
        \justifying
        \footnotetext{This figure shows the average share of employed workers in the formal sector over time for Benin and several comparison countries (Burkina Faso, Cote d'Ivoire, Guinea Bissau, Mali, Niger, Senegal, and Togo). The x-axis represents pre-reform periods (–6 to –1) and post-reform periods (0 to 6), based on the month of survey interview. The y-axis shows the share of employed workers. Solid lines represent individual country trends; the dotted line represents the comparison group mean. Although the LSMS data are cross-sectional, each wave was collected progressively over several months. The monthly event-time variable is constructed using interview dates, with month –6 corresponding to January 2016 and month 0 to January 2019. Individuals are not tracked over time. The sample includes working-age individuals (15–64) who were either employed or unemployed and actively seeking work at the time of the survey. }
        }
    \end{minipage}
\end{figure}

\begin{figure}[H]
    \centering
    \caption{Unadjusted Trend: Share of Employed Workers in Informal Sector}
    \label{mean_employment_informal}
    \includegraphics[width=0.5\textwidth]{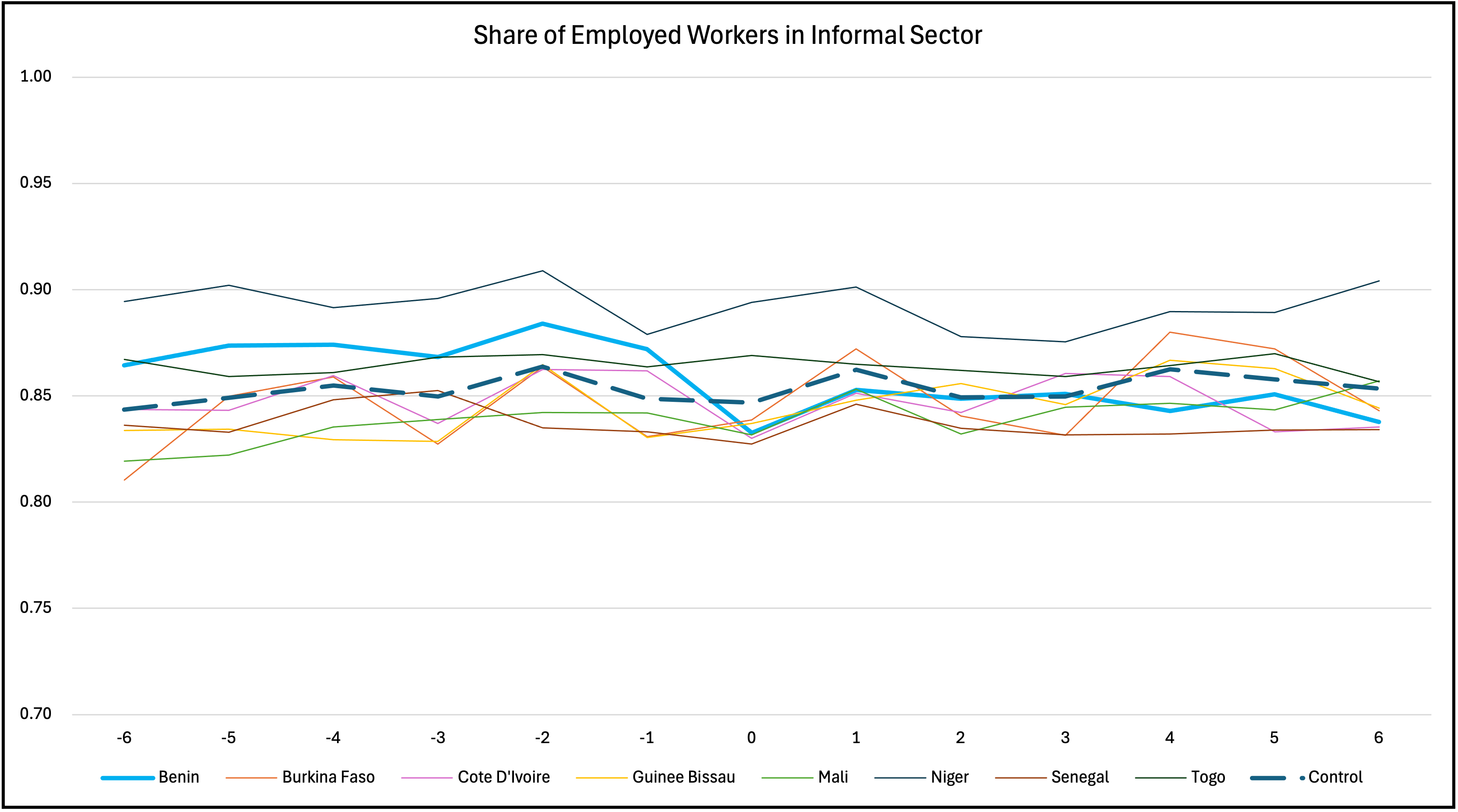}
    \vspace{0.5cm} 
    \begin{minipage}{\textwidth}
        \footnotesize{
        \justifying
        \footnotetext{This figure shows the average share of employed workers in the informal sector over time for Benin and several comparison countries (Burkina Faso, Cote d'Ivoire, Guinea Bissau, Mali, Niger, Senegal, and Togo). The x-axis represents pre-reform periods (–6 to –1) and post-reform periods (0 to 6), based on the month of survey interview. The y-axis shows the share of employed workers. Solid lines represent individual country trends; the dotted line represents the comparison group mean. Although the LSMS data are cross-sectional, each wave was collected progressively over several months. The monthly event-time variable is constructed using interview dates, with month –6 corresponding to January 2016 and month 0 to January 2019. Individuals are not tracked over time. The sample includes working-age individuals (15–64) who were either employed or unemployed and actively seeking work at the time of the survey.}
        }
    \end{minipage}
\end{figure}

The unadjusted trends, shown in Figures \ref{mean_employment}, \ref{mean_employment_formal}, and \ref{mean_employment_informal}, plot the average employment rates over time separately for Benin and the comparison group, distinguishing between total, formal, and informal employment. These figures are constructed using repeated cross-sections to visualize whether pre-reform trends evolved similarly across groups. The visual evidence suggests that pre-reform employment patterns were relatively stable and parallel in both Benin and the control countries, supporting the plausibility of the parallel trends assumption required for the two-way fixed effects strategy. Post-reform, Benin experienced an upward shift in formal employment and a decline in informal employment—patterns not observed in the control group—which is consistent with a causal effect of the reform, though such patterns alone cannot establish causality.

\begin{figure}[H]
    \centering
    \caption{Workers Distribution Across Sectors in Benin}
    \label{sector_distribution}
    \includegraphics[width=\textwidth]{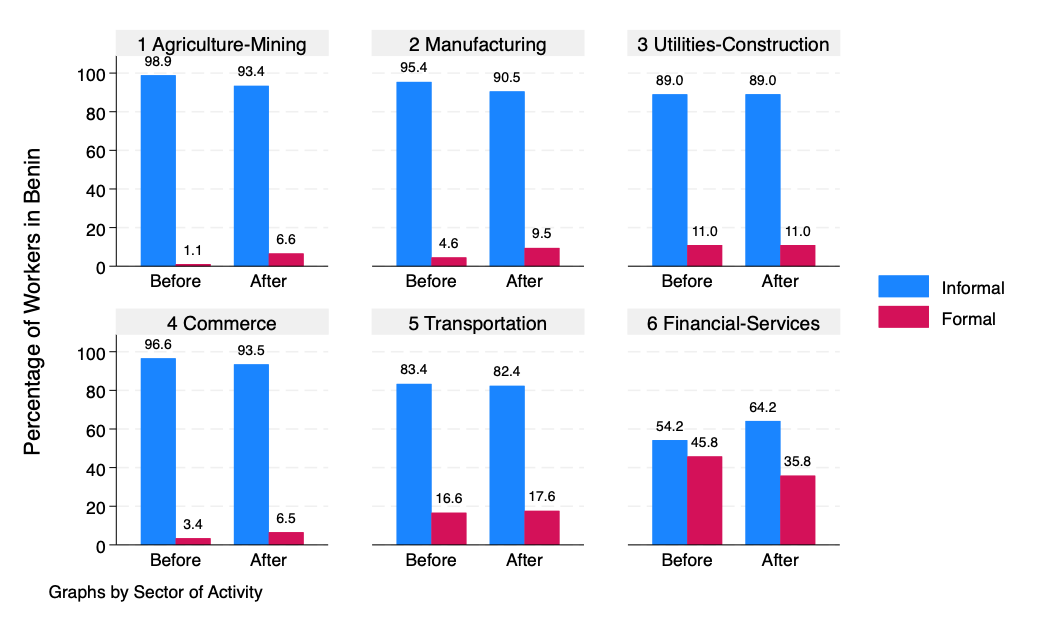}
    \vspace{0.5cm} 
    \begin{minipage}{\textwidth}
        \footnotesize{
        \justifying
        \footnotetext{This figure shows the distribution of workers across different economic sectors in Benin before and after the labor market reform. The x-axis displays two time periods labeled "Before" and "After" the reform. The y-axis represents the percentage of workers in each sector. Blue bars indicate informal employment, while red bars represent formal employment. Each subpanel corresponds to a specific industry: Agriculture-Mining, Manufacturing, Utilities-Construction, Commerce, Transportation, and Financial Services. The most notable shifts include an increase in formal employment within Agriculture-Mining, Manufacturing and Commerce sectors, and moderate gains in formal employment in Transportation. This distribution provides insight into the sector-specific impacts of the reform on formal and informal employment in Benin. Source: By Author.}
        }
    \end{minipage}
\end{figure}

Figure \ref{sector_distribution} illustrates sectoral changes, with a redistribution of labor from traditional industries, such as agriculture, to growing sectors like commerce and financial services. Formal sector growth was concentrated in industries like manufacturing, while informal sector employment remained dominant in agriculture but showed slight declines.

These findings align with existing literature. For instance, \citet{kucera2008informal} provide a comprehensive survey of empirical studies across various developing countries, examining the relationship between formal labor regulations and informal employment. Similarly, \citet{bosch2010comparative} analyze labor market reforms in Brazil and find that such reforms had moderate effects on formal employment. However, Benin’s more pronounced shift suggests that addressing high firing costs and introducing labor market flexibility through STCs had a stronger impact on reducing informality compared to other developing regions. This contrasts with Latin America, where reforms that introduced flexibility often failed to significantly reduce informal employment, as noted by \citet{maloney2004informality} and \citet{levy2008good}.

Overall, while the reform did not drastically change the total employment rate, it successfully redistributed workers between formal and informal sectors, promoting formalization without net job loss. The magnitude of the effect is substantial relative to the size of the formal sector—representing a 24.5\% increase from its pre-reform level—but more moderate in absolute terms relative to the informal sector, which remains dominant. This supports \citet{blanchard2001unemployment}'s theory that labor reforms often lead to labor reallocation rather than significant job creation or destruction.

In conclusion, the 2017 labor market reform in Benin reallocated workers from the informal to the formal sector, reducing informality and promoting employment in more regulated industries. Although the informal sector remains the primary source of employment, the sectoral shifts and stability in overall employment highlight the success of the reform in addressing structural barriers, offering important insights for other countries seeking to implement similar labor market changes.


\subsection{Long-Term Contracts in Formal Sector}

Labor market flexibility reforms can influence the likelihood of workers securing permanent contracts. Reducing firing costs often makes it easier for employers to terminate employees, potentially leading to higher job turnover and a greater proportion of temporary employment arrangements. Employers may prefer temporary contracts, which offer more flexibility and involve fewer long-term commitments. As a result, workers may face reduced opportunities for securing permanent contracts, as employers prioritize short-term arrangements to mitigate economic risks and adapt more easily to market conditions.

As shown in Table \ref{long_term_contract}, the conditional probability of obtaining a permanent contract in the formal sector increased by 23.2 percentage points (SE = 0.020), which represents a 41.6\% increase relative to the pre-reform mean of 55.7\%. This effect captures both the direct impact of the reform on contract conversion within the formal sector and potential composition effects, as the pool of formal sector workers may have changed post-reform. This indicates that, following the reform, a significantly larger proportion of formal sector workers secured long-term contracts, despite the increased flexibility granted to employers. The unconditional probability of obtaining a permanent contract, which accounts for all workers regardless of sector, rose by 4.6 percentage points (SE = 0.004), representing an 88.4\% increase relative to the pre-reform mean of 5.2\%. This broad-based increase suggests that while the reform allowed for more short-term contracts, it also enabled a substantial portion of the workforce to benefit from secure, long-term employment arrangements.

\begin{figure}[H]
    \centering
    \caption{Event Study: Probability of Having a Permanent Contract for workers in the formal sector (Conditional)}
    \label{LTC_cond}
    \includegraphics[width=0.5\textwidth]{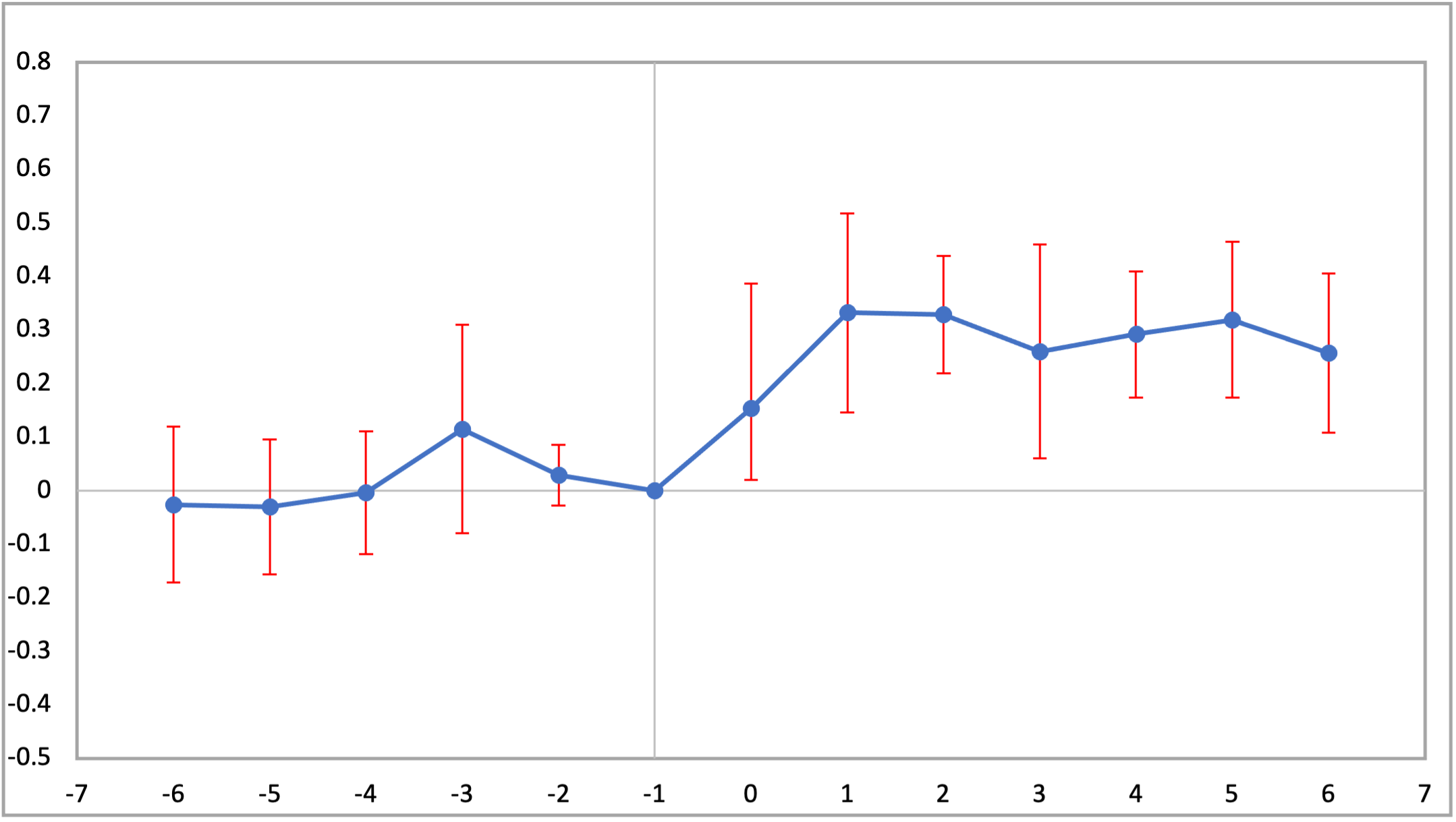}
    \vspace{0.5cm} 
    \begin{minipage}{\textwidth}
        \footnotesize{
        \justifying
        \footnotetext{This event study illustrates the conditional probability of having a long-term contract for workers in the formal sector. The horizontal axis displays periods relative to the reform implementation: lead periods from –6 to –1 represent the months of data collection before the reform, with –1 as the reference period. Periods from 0 to 6 are lags, indicating the months following the reform. Observations show the change in the probability of obtaining a long-term contract post-reform, relative to the pre-reform baseline. Although the LSMS data are cross-sectional, each wave was collected progressively over several months. The monthly event-time variable is constructed based on the interview date, with month –6 corresponding to January 2016 and month 0 to January 2019. Individuals are not tracked over time. The sample includes working-age individuals (15–64) who were either employed or unemployed and actively seeking work at the time of the survey.}
        }
    \end{minipage}
\end{figure}

\begin{figure}[H]
    \centering
    \caption{Event Study: Probability of Having a Permanent Contract (Unconditional)}
    \label{LTC_uncond}
    \includegraphics[width=0.5\textwidth]{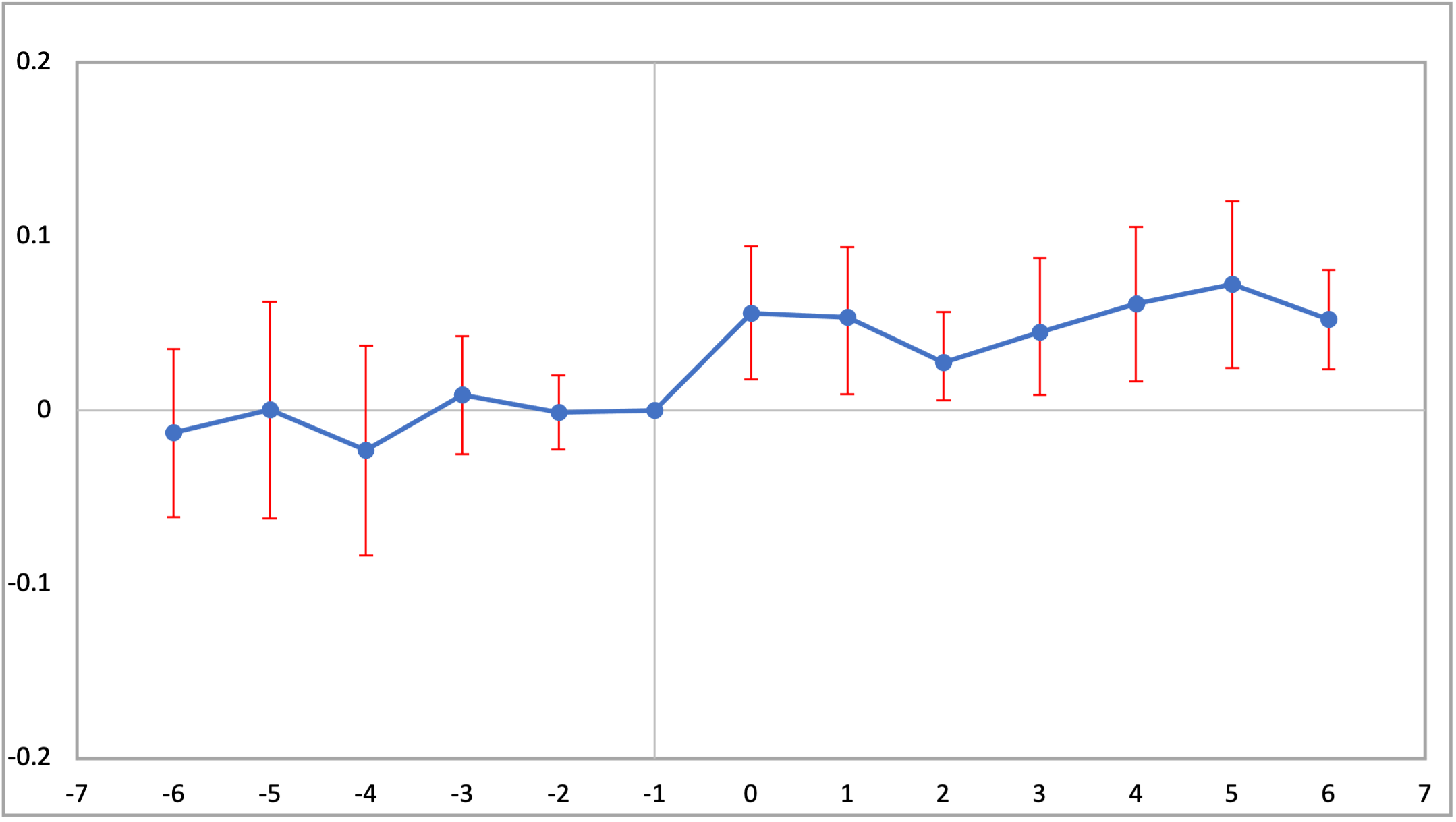}
    \vspace{0.5cm} 
    \begin{minipage}{\textwidth}
        \footnotesize{
        \justifying
        \footnotetext{This event study presents the unconditional probability of having a long-term contract for all workers, without restriction to those in the formal sector. The horizontal axis shows periods relative to the reform: –6 to –1 are lead periods based on the months of data collection prior to the reform, with –1 as the reference period. Periods from 0 to 6 are lags representing the post-reform months. The figure illustrates changes in the probability of obtaining a long-term contract following the reform, relative to the pre-reform baseline. Although the LSMS data are cross-sectional, each wave was collected progressively over several months. The monthly event-time variable is constructed from the interview date, with month –6 corresponding to January 2016 and month 0 to January 2019. Individuals are not tracked over time. The sample includes working-age individuals (15–64) who were either employed or unemployed and actively seeking work at the time of the survey.}
        }
    \end{minipage}
\end{figure}

The event study graphs in Figures \ref{LTC_cond} and \ref{LTC_uncond} provide further insight into these results. The conditional probability of obtaining a long-term contract in the formal sector rose sharply after the reform, as depicted in Figure \ref{LTC_cond}. Before the reform, the probability remained stable, supporting the parallel trends assumption. However, post-reform, there was a marked increase, which remained elevated throughout the subsequent periods. Similarly, Figure \ref{LTC_uncond} shows the unconditional probability for all workers, which also rose after the reform—an increase that is almost mechanical, driven largely by the expansion of formal employment, where long-term contracts are more prevalent. Nonetheless, the simultaneous rise in the conditional probability suggests that firms were not only hiring more workers into the formal sector but were also more likely to offer permanent contracts within that sector. These findings demonstrate that the reform had a wide-reaching impact on long-term employment across the labor market.

\begin{figure}[H]
    \centering
    \caption{Unadjusted Trend: Probability of Having a Permanent Contract in the Formal Sector (Conditional)}
    \label{mean_LTC_cond}
    \includegraphics[width=0.5\textwidth]{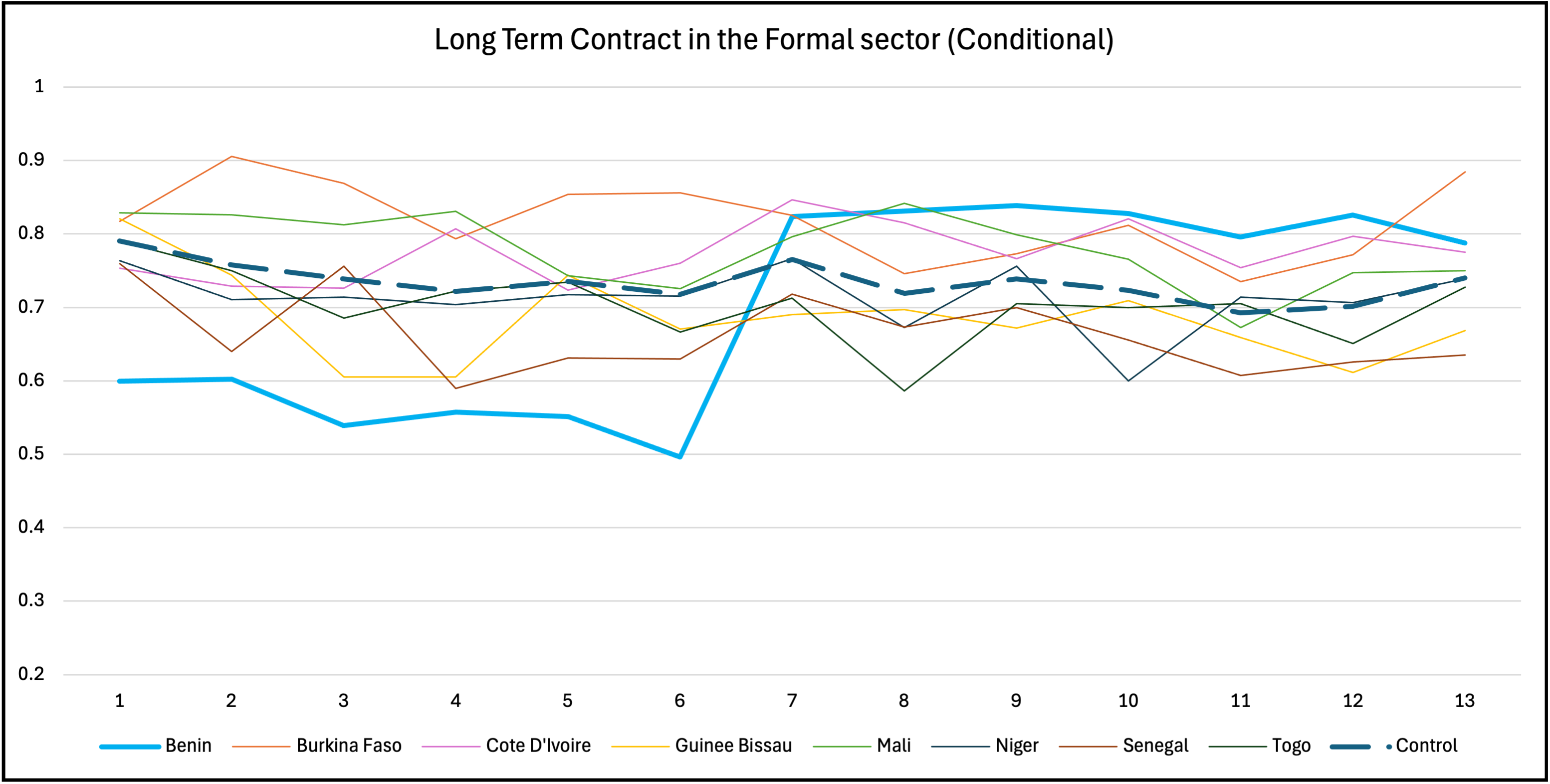}
    \vspace{0.5cm} 
    \begin{minipage}{\textwidth}
        \footnotesize{
        \justifying
        \footnotetext{This figure shows the average share of employed workers who have a permanent contract over time for Benin and several comparison countries (Burkina Faso, Cote d'Ivoire, Guinea Bissau, Mali, Niger, Senegal, and Togo). The x-axis represents time periods, with periods 1 to 6 as pre-reform and 7 to 13 as post-reform. The y-axis shows the share of employed workers. Solid lines represent individual country trends; the dotted line represents the combined mean for the comparison group. Although the LSMS data are cross-sectional, each wave was collected progressively over several months. The monthly event-time variable is based on the interview month, with month –6 corresponding to January 2016 and month 0 to January 2019. Individuals are not tracked over time. The sample includes working-age individuals (15–64) who were either employed or unemployed and actively seeking work at the time of the survey.}
        }
    \end{minipage}
\end{figure}

\begin{figure}[H]
    \centering
    \caption{Unadjusted Trend: Probability of Having a Permanent Contract in the Formal Sector (Unconditional)}
    \label{mean_LTC_uncond}
    \includegraphics[width=0.5\textwidth]{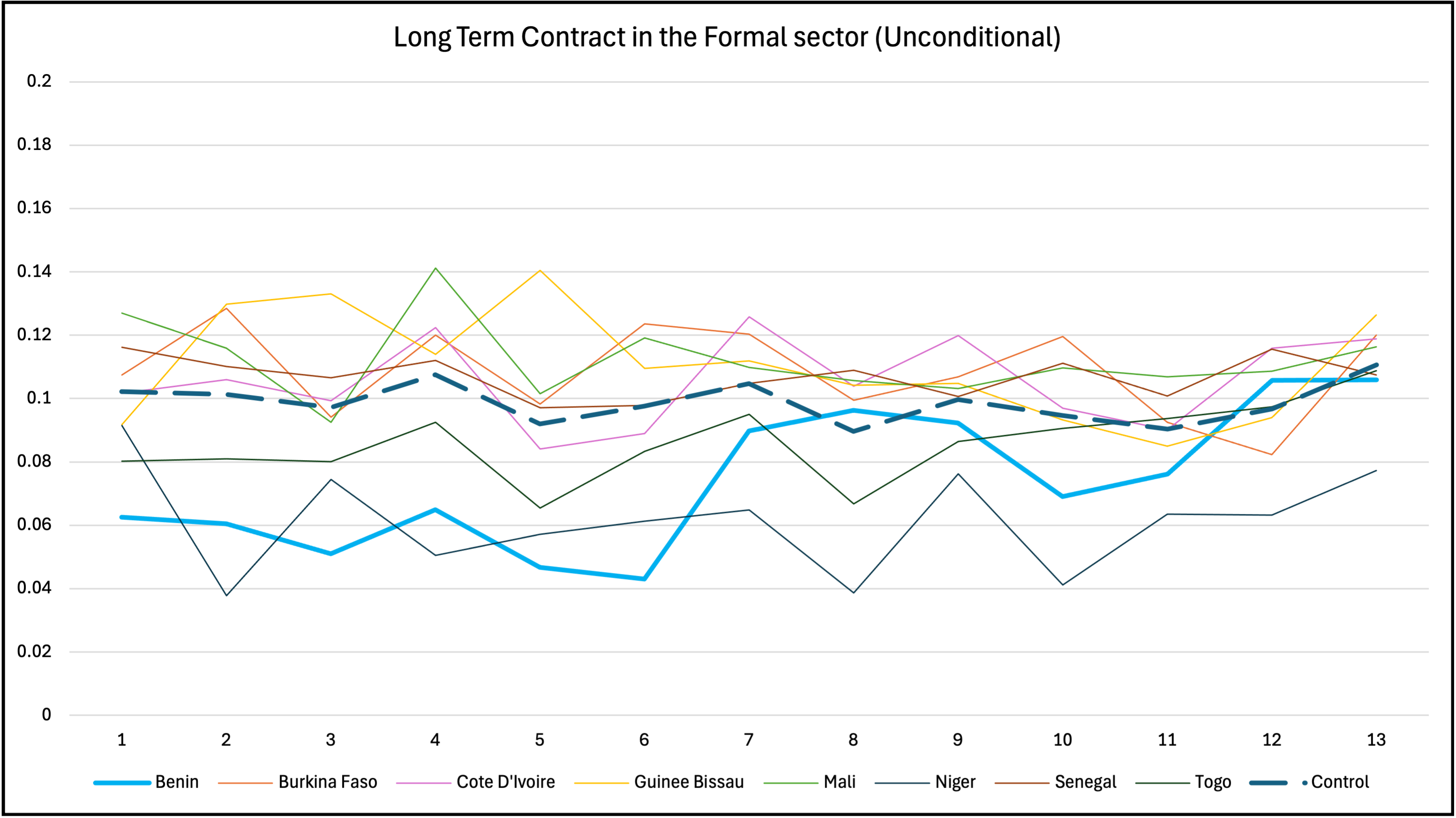}
    \vspace{0.5cm} 
    \begin{minipage}{\textwidth}
        \footnotesize{
        \justifying
        \footnotetext{This figure shows the average share of employed workers who have a permanent contract over time for Benin and several comparison countries (Burkina Faso, Cote d'Ivoire, Guinea Bissau, Mali, Niger, Senegal, and Togo). The x-axis represents time periods, with periods 1 to 6 as pre-reform and 7 to 13 as post-reform. The y-axis shows the share of employed workers. Solid lines represent individual country trends; the dotted line represents the combined mean for the comparison group. Although the LSMS data are cross-sectional, each wave was collected progressively over several months. The monthly event-time variable is based on the interview month, with month –6 corresponding to January 2016 and month 0 to January 2019. Individuals are not tracked over time. The sample includes working-age individuals (15–64) who were either employed or unemployed and actively seeking work at the time of the survey.}
        }
    \end{minipage}
\end{figure}

The unadjusted trends in Figures \ref{mean_LTC_cond} and \ref{mean_LTC_uncond} reinforce these conclusions. The average conditional probability of securing a permanent contract remained stable for Benin before the reform, tracking closely with comparison countries. However, following the reform, Benin experienced a sharp increase in this probability, while the trends in the comparison group remained flat. This divergence illustrates that the observed changes in Benin were driven by the reform itself, underscoring its substantial impact on employment dynamics.

The reform’s effect on the probability of obtaining a permanent contract was not uniform across different demographic groups. Table \ref{long_term_contract_heterogeneity} presents the heterogeneity analysis, revealing that women experienced a larger increase in the conditional probability of securing a long-term contract (28.0 percentage points) compared to men (21.1 percentage points). While the difference between the two is not statistically significant (t-test value = -1.48), the relatively larger gain for women suggests that the reform may have disproportionately benefited female workers, who often face greater labor market insecurity. Rural workers, in particular, saw a more pronounced increase (29.3 percentage points) compared to urban workers (19.0 percentage points), with the t-test value of -2.04 indicating a significant difference. This result suggests that the reform had a stronger impact in rural areas, possibly reflecting greater labor market competition in these regions, where employers may have needed to offer more permanent contracts to retain workers.

Age also played a role in the reform’s impact, with workers under 35 seeing a slightly larger increase in the probability of securing a permanent contract (25.1 percentage points) compared to those over 35 (19.3 percentage points), though the difference was not statistically significant (t-test value = 1.55). Younger workers may have been more likely to benefit from the reform as they entered the labor market under more flexible conditions, enabling them to secure longer-term employment arrangements earlier in their careers. Marital status did not lead to significant differences, as both married and non-married workers experienced similar increases in the probability of obtaining a permanent contract, with t-test values suggesting no meaningful distinction between the two groups.

\begin{figure}[H]
    \centering
    \caption{Tenure Histograms: Formal and Informal Sectors Before and After the Reform}
    \label{tenure_histogram}
    \begin{minipage}{0.49\textwidth}
        \centering
        \includegraphics[width=\textwidth]{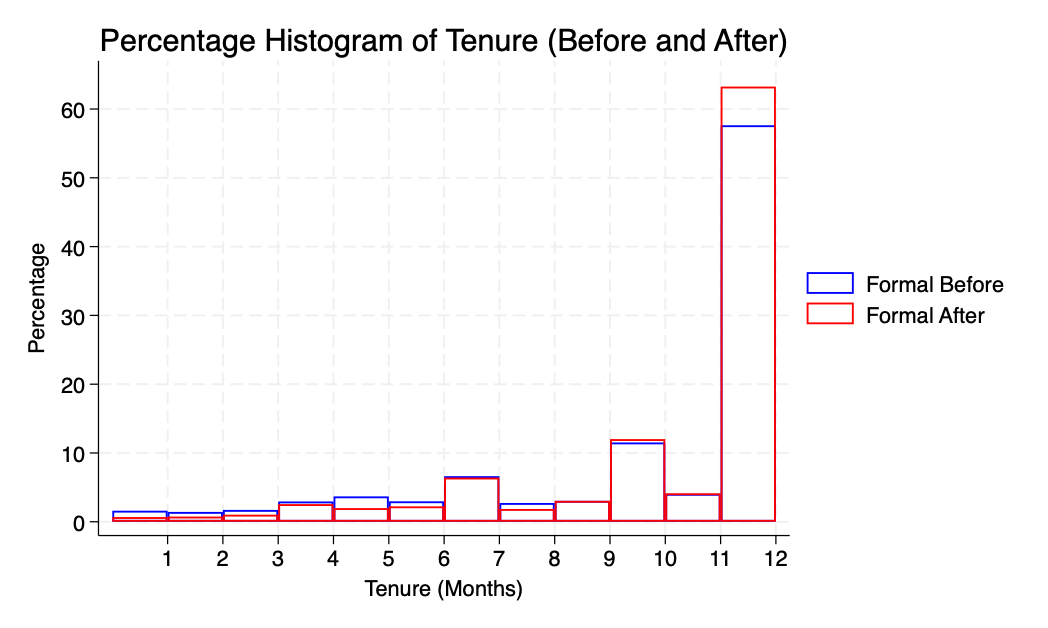}
        \subcaption{Formal Sector}
    \end{minipage}%
    \hspace{0.1cm} 
    \begin{minipage}{0.49\textwidth}
        \centering
        \includegraphics[width=\textwidth]{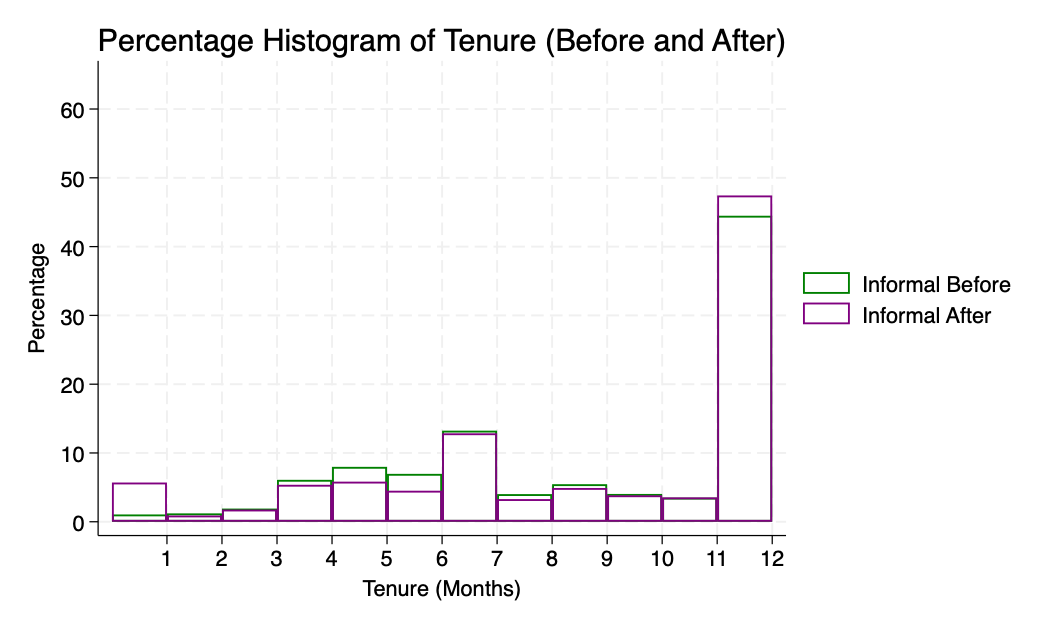}
        \subcaption{Informal Sector}
    \end{minipage}
    \vspace{0.5cm} 
    \begin{minipage}{\textwidth}
        \footnotesize{
        \justifying
        \footnotetext{
        The histograms display the distribution of job tenure (in months) among workers in the formal and informal sectors before and after the reform. In the formal sector (left panel), the post-reform distribution (red) shows a notable increase in the percentage of workers with 12 months of tenure, indicating greater retention or job stability following the reform. The informal sector (right panel) also exhibits a shift in tenure distribution, with more workers reaching higher tenure levels (shown in purple) after the reform compared to the pre-reform period (green). These patterns suggest that the reform may have influenced both formal and informal sector job tenure. Source: By Author.
        }
        }
    \end{minipage}
\end{figure}

Figure \ref{tenure_histogram} provides additional context by displaying the distribution of job tenure in both the formal and informal sectors before and after the reform. In the formal sector, the post-reform histogram shows a marked increase in the percentage of workers with 12 months of tenure, reflecting greater job stability and retention. The informal sector also experienced an upward shift, though the effect was less pronounced compared to the formal sector. This indicates that the reform’s effects on employment stability extended beyond the formal sector, potentially creating spillover effects that improved job tenure even for informal workers.

The increase in permanent contracts also suggests that firms valued long-term employment relationships to retain skilled workers and improve productivity. In sectors where long-term contracts foster worker loyalty or efficiency, firms may have opted to offer more permanent contracts despite the increased labor market flexibility introduced by the reform. This pattern is particularly noteworthy in a developing country like Benin, where informal employment is widespread, and long-term contracts have traditionally been rare. The findings challenge the commonly held view that labor market flexibility reforms primarily lead to precarious employment. While temporary contracts undoubtedly provided employers with flexibility, the evidence from Benin suggests that well-designed reforms can promote more stable employment arrangements. This aligns with theories suggesting that firms, particularly in labor markets with significant skill demands, still find value in offering permanent contracts even in a more flexible regulatory environment.

In conclusion, the 2017 labor market reform in Benin led to a significant increase in both the conditional and unconditional probabilities of obtaining a long-term contract. The event study and unadjusted trends confirm that the reform drove these changes, while the heterogeneity analysis highlights that women, rural workers, and younger employees benefited the most. These findings suggest that the reform successfully balanced labor market flexibility with job stability, creating opportunities for long-term employment across a broad section of the workforce. By promoting long-term contracts in conjunction with increased flexibility, the reform has important implications for other developing economies seeking to implement similar labor market changes.


\subsection{Tenure (Last 12 Months)}

The 2017 labor market reform in Benin had a noticeable impact on the employment tenure of workers, revealing changes in the stability of jobs across different sectors and contract types. Tenure is a key indicator of job stability, representing the number of months workers have stayed in their current positions over the last 12 months.\footnote{Although tenure reflects the previous 12 months, the 2019 survey was conducted well after the August 2017 reform, ensuring that the tenure data fully capture post-reform employment conditions.
} This outcome is truncated above at 12 months. The analysis of tenure, both conditional and unconditional, across the formal and informal sectors provides valuable insights into how the reform influenced employment duration.

As shown in Table \ref{tenure_12_months}, the reform led to an overall reduction in worker tenure by 0.158 months (approximately 4.74 days, SE = 0.059), indicating that while the reform may have increased flexibility for employers, it also slightly decreased the duration workers remained in their jobs. This reduction represents a 1.66\% decrease relative to the mean pre-reform tenure of 9.533 months, suggesting that greater labor market flexibility came at the cost of shorter job durations, particularly for those in short-term contracts.

When examining the formal sector, the results show different effects depending on whether short-term contract (STC) workers are included. The conditional tenure for all workers in the formal sector, reported in column (2), decreased slightly by 0.118 months (SE = 0.108), but this reduction was not statistically significant. This implies that the tenure of formal sector workers remained relatively stable overall. However, the unconditional tenure for formal sector workers, which includes those not employed in the formal sector as having zero tenure, increased significantly by 0.153 months (SE = 0.039), a 15.45\% rise relative to the pre-reform mean of 0.990 months. This increase suggests that while more workers may have entered the formal sector under short-term contracts, some of them secured longer job durations.

For formal sector workers on short-term contracts (STC), the reform had a mixed effect. Conditional tenure for STC workers, shown in column (4), fell by 0.070 months (SE = 0.231), though this reduction was not statistically significant. However, the unconditional tenure for STC workers in the formal sector dropped significantly by 0.229 months (SE = 0.028), or 50.33\%, relative to the pre-reform mean of 0.455 months. This indicates that the reform increased labor market flexibility by expanding the use of short-term contracts, but it also shortened job durations for workers on these contracts, leading to more frequent turnover.

In the informal sector, the results are more pronounced. Conditional tenure for informal sector workers under STCs decreased by 0.148 months (SE = 0.060), a significant reduction of 1.55\% relative to the pre-reform mean of 9.556 months. Unconditional tenure in the informal sector also declined significantly by 0.311 months (SE = 0.068), a 3.64\% drop from the pre-reform mean of 8.543 months. These reductions suggest that the reform, while focusing on the formal sector, may have led to shorter job durations in the informal sector as well. Workers in the informal sector may have faced increased competition from formal sector opportunities, or informal employers may have adopted more flexible employment practices similar to those introduced in the formal sector.

\begin{figure}[H]
    \centering
    \caption{Event Study: Tenure Last 12 Months (overall)}
    \label{edd_tenure_overall}
    \includegraphics[width=0.5\textwidth]{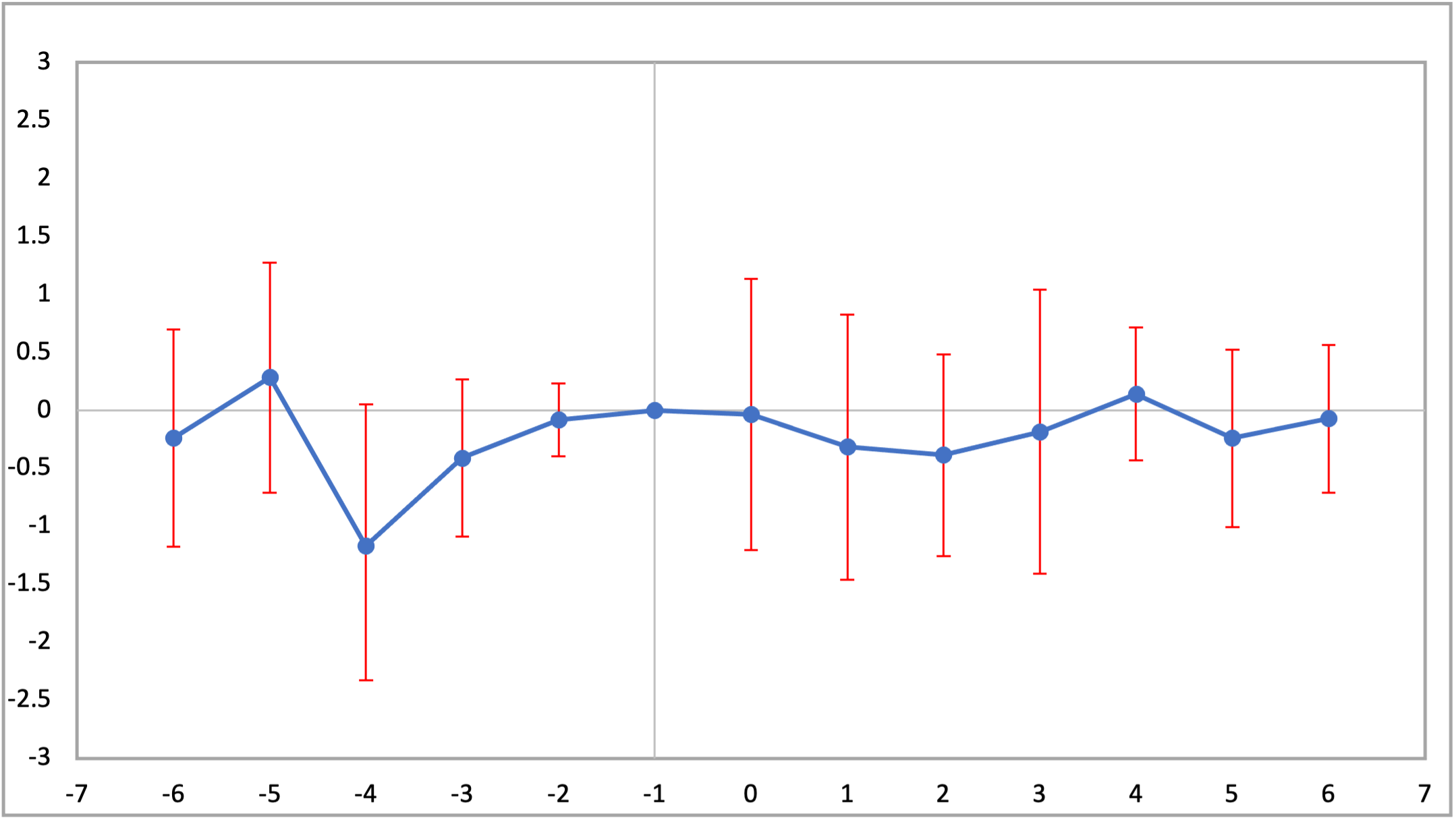}
    \vspace{0.5cm} 
    \begin{minipage}{\textwidth}
        \footnotesize{
        \justifying
        \footnotetext{
        This event study illustrates the average tenure (in months) over the last 12 months for workers across all sectors, without restriction to formal or informal employment (unconditional tenure). The x-axis represents time periods relative to the reform implementation, where –6 to –1 are lead periods, indicating months of data collection before the reform, with –1 as the reference period. Periods from 0 to 6 are lags, representing months after the reform. The y-axis shows the deviation in tenure from the reference period, allowing us to observe changes in tenure trends post-reform compared to the pre-reform baseline. Although the LSMS data are cross-sectional, each wave was collected progressively over several months. The monthly event-time variable is constructed based on the interview month, with month –6 corresponding to January 2016 and month 0 to January 2019. Individuals are not tracked over time. The sample includes working-age individuals (15–64) who were either employed or unemployed and actively seeking work at the time of the survey.
        }
        }
    \end{minipage}
\end{figure}

The event study in Figure \ref{edd_tenure_overall} illustrates the overall changes in tenure trends before and after the reform. In the pre-reform period (lead periods from -6 to -1), tenure remained relatively stable across all sectors. However, following the reform (periods from 0 to 6), there was a noticeable decline in tenure, particularly in the informal sector. This indicates that the reform had an immediate impact on reducing job duration, likely due to the increased prevalence of short-term contracts and greater employer flexibility in hiring and firing decisions.

\begin{figure}[H]
    \centering
    \caption{Unadjusted Trends: Tenure Last 12 Months (Unconditional)}
    \label{tenure_unconditional}
    \includegraphics[width=0.5\textwidth]{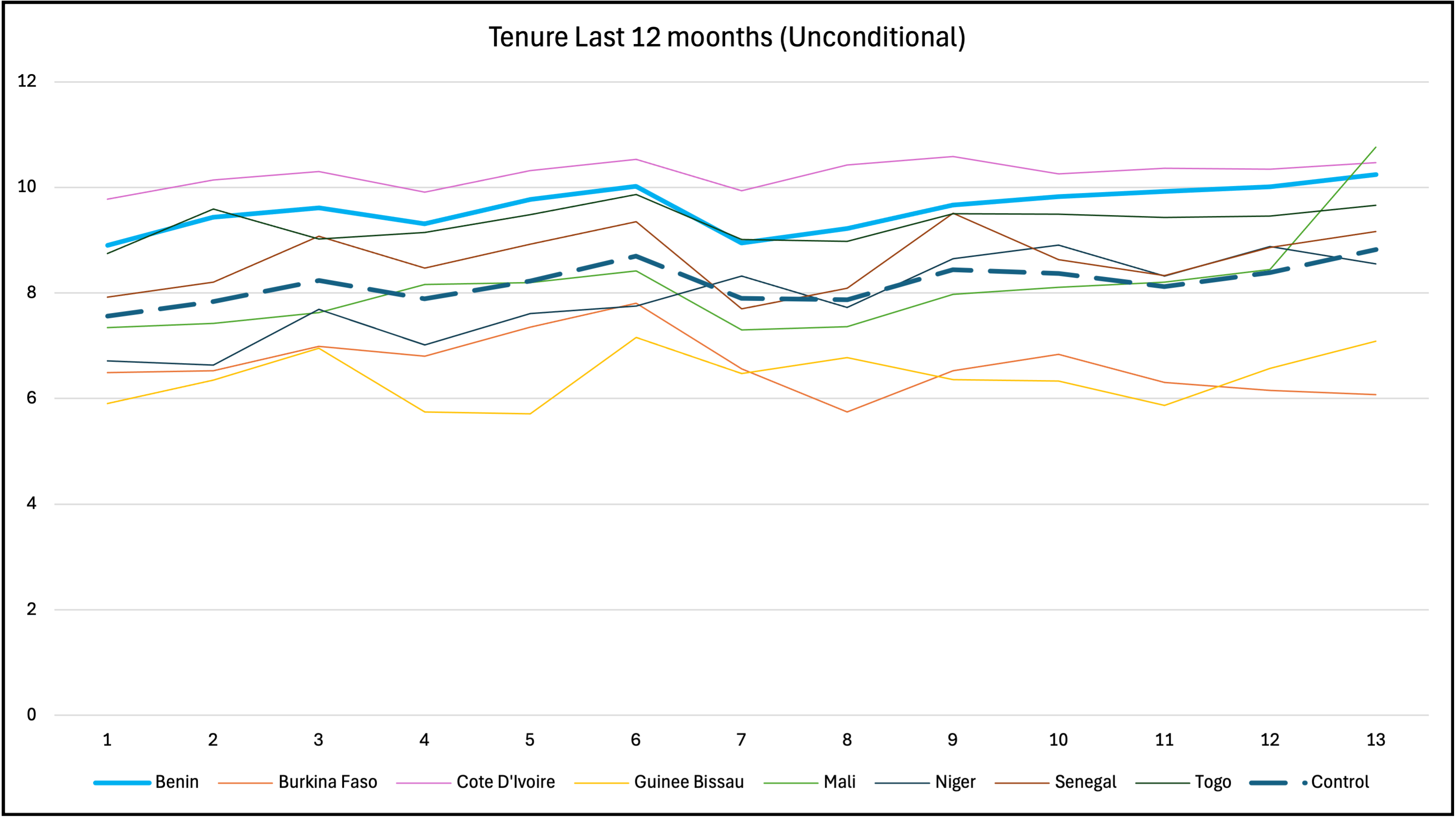}
    \vspace{0.5cm} 
    \begin{minipage}{\textwidth}
        \footnotesize{
        \justifying
        \footnotetext{This figure displays the average tenure (in months) over the last 12 months for workers across different countries, irrespective of their sector of employment (unconditional tenure). Each line represents a country, with Benin indicated by a solid cyan line. The control group average is shown with a dashed line. The x-axis represents time periods, with periods 1 to 6 as pre-reform and 7 to 13 as post-reform. The y-axis shows the average tenure. This visualization compares tenure trends across countries—Benin, Burkina Faso, Cote d'Ivoire, Guinea Bissau, Mali, Niger, Senegal, and Togo—both individually and relative to the control. Although the LSMS data are cross-sectional, each wave was collected progressively over several months. The monthly event-time variable is based on the interview month, with month –6 corresponding to January 2016 and month 0 to January 2019. Individuals are not tracked over time. The sample includes working-age individuals (15–64) who were either employed or unemployed and actively seeking work at the time of the survey.}
        }
    \end{minipage}
\end{figure}

The unadjusted trends in Figure \ref{tenure_unconditional} provide further context. In this figure, the tenure trends across multiple countries are compared, with Benin indicated by a solid cyan line. The control group, shown by the dashed line, experienced relatively stable tenure throughout the period. However, in Benin, there is a clear downward shift in tenure after the reform, particularly when compared to the other countries in the control group. This decline may partially reflect a compositional effect, as the reform brought a wave of new hires into the formal sector—individuals who, by construction, have shorter tenure at the time of survey. This visual evidence supports the regression results, demonstrating that the reform significantly reduced job durations in Benin, while other West African countries did not experience similar declines in tenure during the same period.

The histograms in Figure \ref{tenure_histogram} further illustrate the distribution of tenure across the formal and informal sectors before and after the reform. In the formal sector (left panel), the post-reform distribution shows a clear increase in the percentage of workers with exactly 12 months of tenure, suggesting that more workers were employed for a full year. However, there is a noticeable decline in mid-range tenure (1 to 11 months), indicating that while some workers reached a full year of employment, others experienced shorter job spells. In the informal sector (right panel), the post-reform distribution shows a broader shift toward shorter tenure, with a larger proportion of workers reporting shorter employment spells. This further supports the conclusion that the reform led to increased turnover and shorter job durations, particularly in the informal sector.

In conclusion, the 2017 labor market reform in Benin had a mixed impact on employment tenure. While the reform created new job opportunities and increased flexibility, it also led to shorter job durations across both formal and informal sectors. Workers on short-term contracts experienced the most significant reductions in tenure, while those in the informal sector saw job stability decline. The overall reduction in tenure reflects the increased flexibility introduced by the reform, but it also raises concerns about the long-term stability of jobs in a more flexible labor market.


\subsection{Non-Employment Spell in Years}

The 2017 labor market reform in Benin had notable effects on non-employment spells, which measure how long individuals spent without employment. This indicator, expressed in years\footnote{This measure is based on self-reported time since last job and may be subject to rounding or recall imprecision, especially when the duration is reported in approximate months. Nevertheless, the LSMS question provides a direct and widely used indicator of non-employment duration.}, helps us understand whether the reform contributed to extended periods of non-employment for workers. The analysis, both conditional (restricted to currently non-employed individuals) and unconditional (across the entire sample), provides insight into the dynamics of non-employment and how they shifted after the reform.

Table \ref{unemployment_spell} shows that the conditional effect—focusing solely on workers who were non-employed—suggests that the reform led to an increase in the average non-employment spell by 0.397 years (SE = 0.228). This represents a 29.19\% rise from the pre-reform average of 1.360 years, indicating that, for those who experienced non-employment, the reform may have prolonged their time out of work. However, the confidence interval for this estimate includes zero (95\% CI: [-0.051, 0.846]), so the result is not statistically significant. This extended duration may reflect challenges for workers in finding stable employment, possibly due to an increase in short-term contracts, which, while offering employment opportunities, may have also led to more frequent gaps between jobs.

In contrast, the unconditional effect, which considers the entire sample (both employed and non-employed), is much smaller. The reform is associated with a modest increase in non-employment duration by 0.010 years (SE = 0.005), or 62.5\% above the pre-reform mean of 0.016 years. While this effect is also not statistically significant (95\% CI: [-0.001, 0.020]), it suggests that for the broader population, including those who remained employed, the reform did not result in a large shift in the overall pattern of non-employment spells.

\begin{figure}[H]
    \centering
    \caption{Event Study: Non-Employment Spell in Years (Conditional)}
    \label{unem_spell_cond}
    \includegraphics[width=0.5\textwidth]{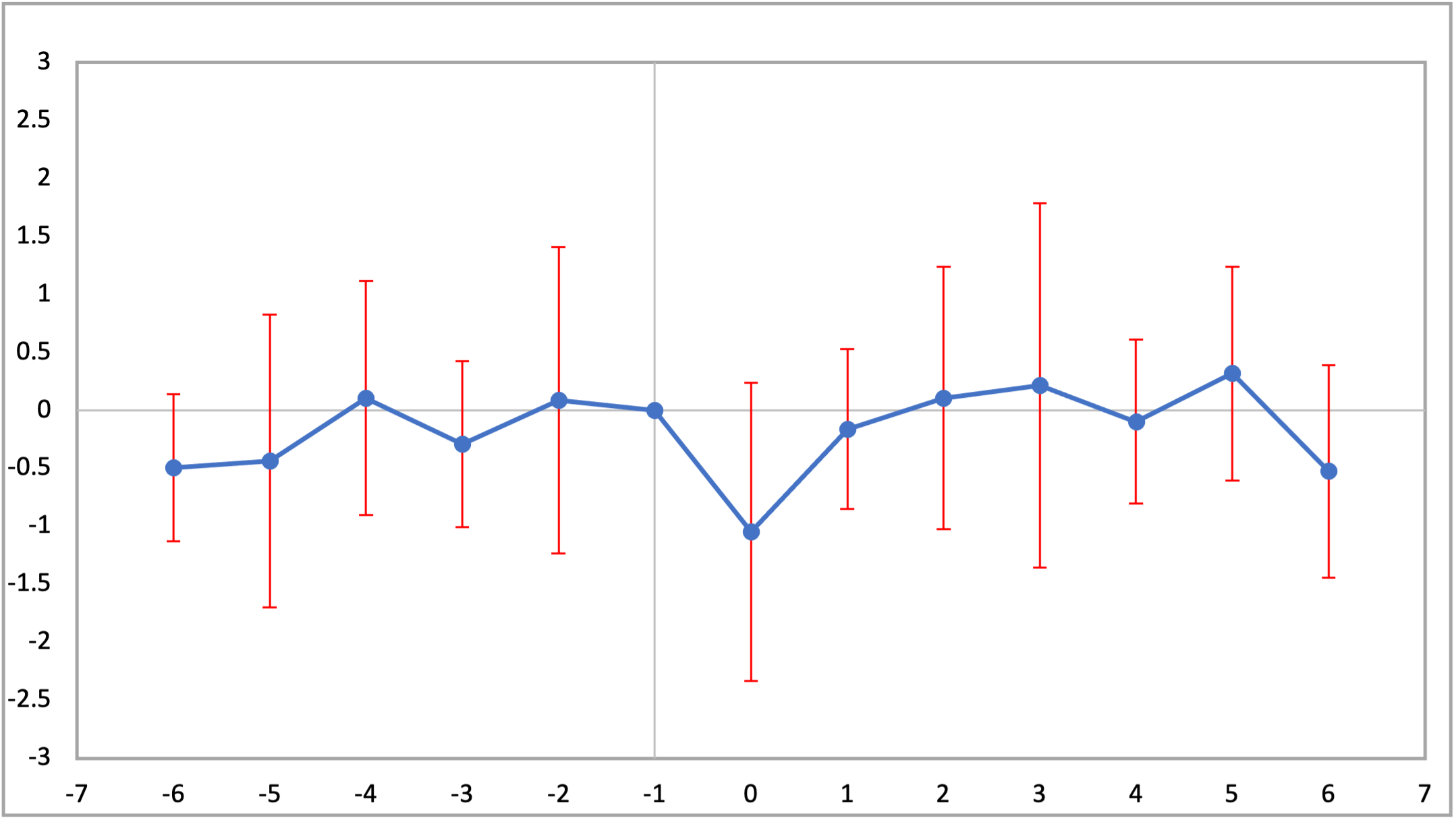}
    \vspace{0.5cm} 
    \begin{minipage}{\textwidth}
        \footnotesize{
        \justifying
        \footnotetext{
        This event study illustrates the average duration of unemployment spells in years for workers who are currently unemployed (conditional sample). The x-axis represents time periods relative to the reform implementation, where –6 to –1 are lead periods indicating months of data collection before the reform, with –1 serving as the reference period. Periods from 0 to 6 are lags, representing months of data collection following the reform. The y-axis shows the deviation in unemployment spell duration from the reference period, allowing us to observe changes in unemployment duration trends post-reform compared to the pre-reform baseline. Although the LSMS data are cross-sectional, each wave was collected progressively over several months. The monthly event-time variable is constructed based on the interview date, with month –6 corresponding to January 2016 and month 0 to January 2019. Individuals are not tracked over time. The sample includes working-age individuals (15–64) who were either employed or unemployed and actively seeking work at the time of the survey.
        }
        }
    \end{minipage}
\end{figure}

The event study in Figure \ref{unem_spell_cond} visually confirms this trend, showing that non-employment spells for currently non-employed workers remained stable before the reform (in the lead periods from -6 to -1), but there is a slight rise in non-employment durations after the reform (lag periods from 0 to 6). Although the increase is not dramatic, it indicates that the reform may have introduced additional friction for workers seeking to return to employment.


These results point to a nuanced outcome of the labor market reform. For workers who were already out of employment, the reform may have contributed to longer non-employment spells, suggesting potential difficulties in finding or maintaining stable work in a more flexible labor market. The increase in short-term contracts and reduced firing costs may have resulted in more job turnover, leading to frequent gaps between periods of employment for certain workers. However, for the general workforce, the reform did not significantly extend the average duration of non-employment, indicating that its effects were more concentrated among specific groups.

While the reform aimed to enhance labor market flexibility, it is important to consider its potential trade-offs, particularly for those who struggled to find stable work after experiencing non-employment. The findings suggest that while the reform may have facilitated job entry for some, it also may have contributed to longer spells of non-employment for others.


\subsection{Monthly Earnings in the Formal Sector}

The 2017 labor market reform in Benin aimed to reduce firing costs and increase labor market flexibility, with potential implications for earnings in the formal sector. By reducing restrictions on hiring and firing, the reform could have increased short-term employment opportunities while influencing wage dynamics in the formal sector.

Table \ref{wages_formal_sector_did} presents the two way fixed effect estimates for monthly earnings in the formal sector, with results disaggregated for short-term contract (STC) and long-term contract (LTC) workers. Column (1) shows the conditional results for workers currently employed in the formal sector, where the reform led to a significant increase in monthly earnings by 33.615 USD (SE = 10.632), representing a substantial rise relative to the mean pre-treatment wage of 118.602 USD. This finding suggests that for workers in the formal sector, the reform created more lucrative employment opportunities, potentially as a result of employers using the increased flexibility to offer higher wages to retain skilled workers in key positions.

In contrast, when considering the unconditional results (Column 2), which include workers from both the formal and informal sectors (with informal sector workers having zero wages), the reform's impact on monthly earnings is much smaller and statistically insignificant. The estimated effect is only 1.424 USD (SE = 1.258), with a 95\% confidence interval that includes zero [-1.046, 3.894], suggesting that the wage effects of the reform were not widespread across the entire population but concentrated in the formal sector.

Looking specifically at workers with short-term contracts (STC) in the formal sector, Column (3) shows a more modest increase in monthly earnings by 19.625 USD (SE = 9.668), though this effect is not statistically significant (95\% CI: [-0.607, 38.644]). This result may indicate that while the reform expanded the use of short-term contracts, it did not lead to significantly higher wages for STC workers in the formal sector, likely due to the temporary and flexible nature of these contracts.

For workers with long-term contracts (LTC), as shown in Column (5), the reform is associated with a significant wage increase of 23.411 USD (SE = 11.601). This rise in earnings may reflect that while the reform introduced greater labor market flexibility, firms continued to offer higher wages to LTC workers, possibly to retain valuable employees with long-term commitments. This trend aligns with the broader findings on job stability and the reform's effect on permanent contracts.

\begin{figure}[H]
    \centering
    \caption{Event Study: Monthly Earnings in the formal sector in USD (Conditional)}
    \label{wage_cond}
    \includegraphics[width=0.5\textwidth]{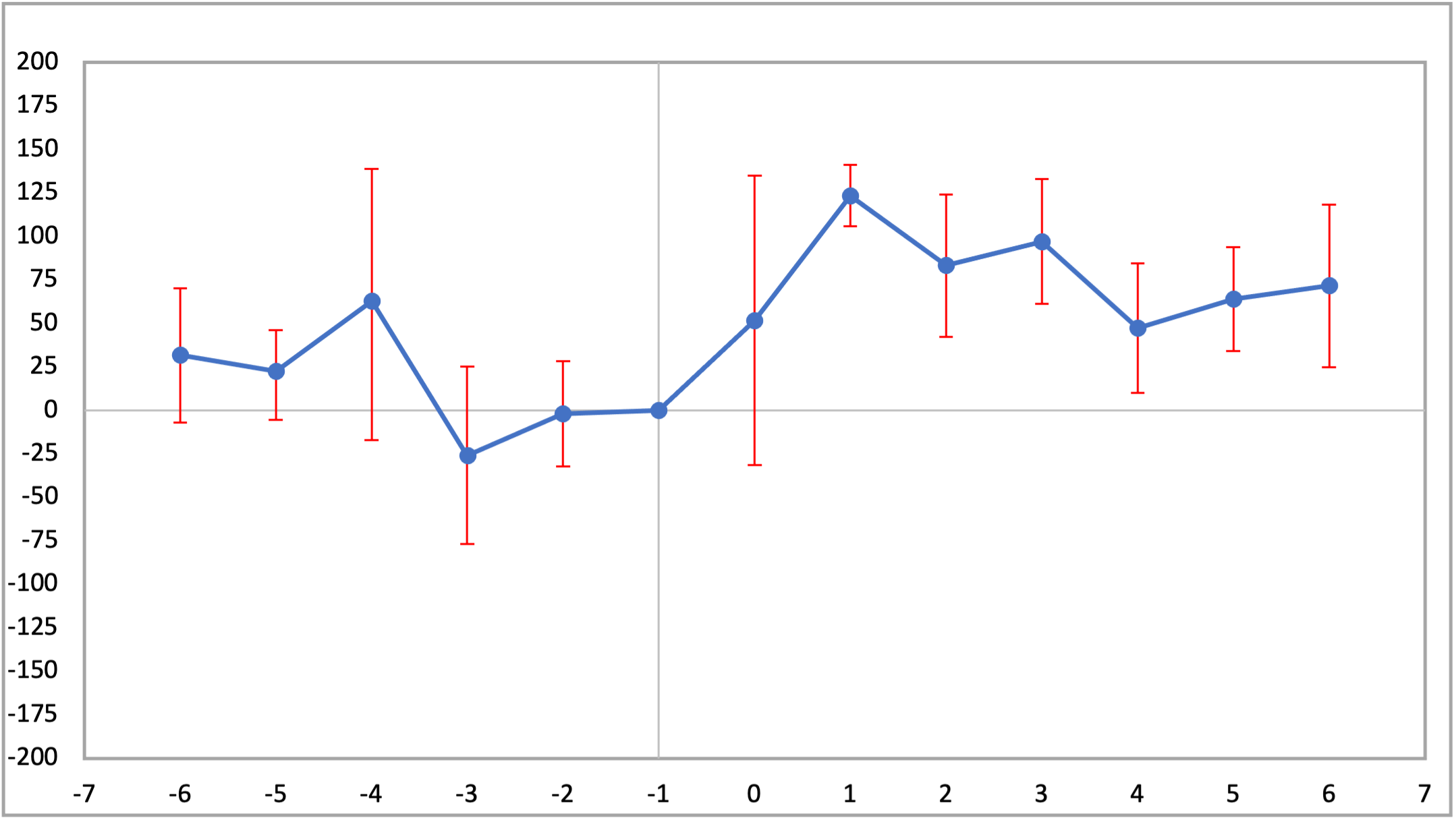}
    \vspace{0.5cm} 
    \begin{minipage}{\textwidth}
        \footnotesize{
        \justifying
        \footnotetext{
        This event study illustrates the average monthly earnings for workers who are currently in the formal sector (conditional sample). The x-axis represents time periods relative to the reform implementation, where –6 to –1 are lead periods indicating months of data collection before the reform, with –1 serving as the reference period. Periods from 0 to 6 are lags, representing months of data collection following the reform. The y-axis shows the deviation in wages from the reference period, allowing us to observe changes in earnings trends post-reform compared to the pre-reform baseline. Although the LSMS data are cross-sectional, each wave was collected progressively over several months. The monthly event-time variable is constructed based on the interview date, with month –6 corresponding to January 2016 and month 0 to January 2019. Individuals are not tracked over time. The sample includes working-age individuals (15–64) who were either employed or unemployed and actively seeking work at the time of the survey.        }
        }
    \end{minipage}
\end{figure}

The event study presented in Figure \ref{wage_cond} illustrates the conditional impact of the reform on wages over time. Before the reform (lead periods from -6 to -1), wages were relatively stable across time periods. However, following the reform (lag periods from 0 to 6), there is a noticeable upward shift in monthly earnings for formal sector workers, confirming the significant positive effect observed in the regression results.

\begin{figure}[H]
    \centering
    \caption{Event Study: Monthly Earnings in the formal sector in USD (Unconditional)}
    \label{wage_uncond}
    \includegraphics[width=0.5\textwidth]{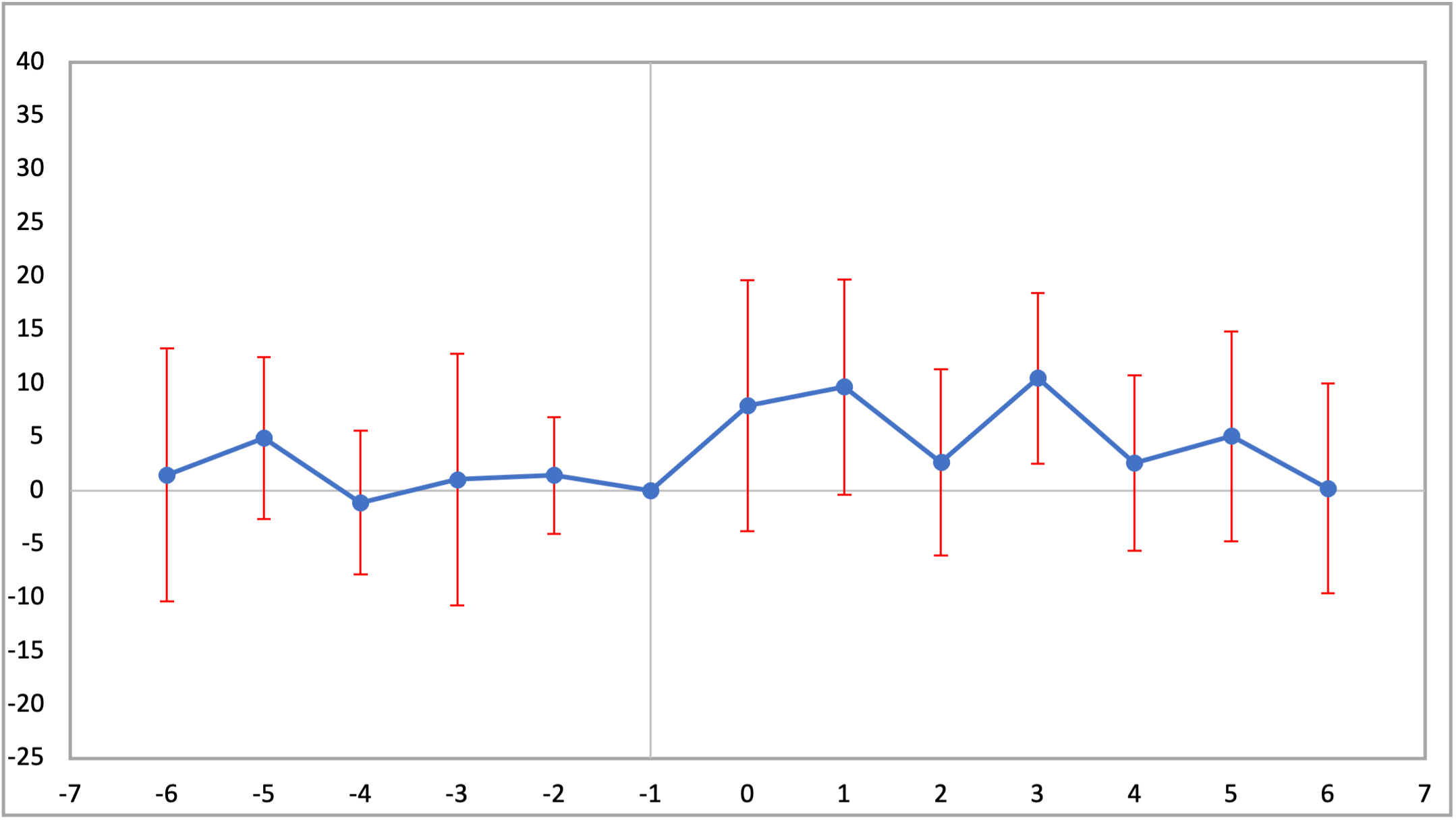}
    \vspace{0.5cm} 
    \begin{minipage}{\textwidth}
        \footnotesize{
        \justifying
        \footnotetext{
        This event study illustrates the average monthly earnings in the formal sector across all workers, without restriction to workers in the formal sector (unconditional sample). The x-axis represents time periods relative to the reform implementation, where –6 to –1 are lead periods, with –1 as the reference period. Periods from 0 to 6 are lags, representing months of data collection post-reform. The y-axis shows the deviation in wages from the reference period, allowing us to observe changes in earnings trends following the reform. Although the LSMS data are cross-sectional, each wave was collected progressively over several months. The monthly event-time variable is constructed based on the interview month, with month –6 corresponding to January 2016 and month 0 to January 2019. Individuals are not tracked over time. The sample includes working-age individuals (15–64) who were either employed or unemployed and actively seeking work at the time of the survey.
        }
        }
    \end{minipage}
\end{figure}

Figure \ref{wage_uncond} shows the unconditional wage trends for all workers, including those in the informal sector (with zero wages). The figure suggests that while the reform led to increases in wages for formal sector workers, the overall impact on wages, when considering the entire population, was modest.

These findings suggest that the labor market reform led to significant wage increases for formal sector workers, particularly those with long-term contracts. However, the reform's effect on wages was less pronounced for short-term contract workers and when considering the entire labor force, indicating that the wage benefits were concentrated within specific segments of the formal sector workforce. It is important to note, however, that part of the observed wage increase—especially among long-term contract holders—may reflect compositional changes in the post-reform workforce, such as a higher share of more educated or productive individuals entering formal jobs. Thus, the increase in wages may result from both changes in firm behavior and shifts in the types of workers employed.


\section{Theoretical Model of Job Search}\label{model}

To better understand the mechanisms underlying the labor market impacts of Benin’s 2017 reform, I develop a job search model based on the Diamond-Mortensen-Pissarides (DMP) framework. The model serves to interpret the empirical findings and assess potential counterfactual scenarios related to changes in contract structures and firm behavior. The DMP framework explains labor market frictions through the coexistence of unemployment and job vacancies in equilibrium. These frictions stem from imperfect information and time-consuming search processes that prevent instantaneous matching, leading to equilibrium outcomes characterized by unemployment, job creation, and endogenous mobility.

While the core structure follows the standard DMP framework, this model incorporates several extensions that reflect the institutional and economic realities of Benin’s labor market. First, the model introduces endogenous job separation, where the decision to end a job match depends on the realization of a match-specific productivity draw. This is in contrast to the typical assumption of constant exogenous separation. Second, the model includes heterogeneity among firms and workers, where match productivity $z \sim F(z)$ varies across matches. This heterogeneity drives sorting and contract transitions. Matches with low productivity realizations are more likely to be dismissed or remain on short-term contracts, while higher-productivity matches are more likely to be upgraded to long-term contracts. The endogenous evaluation of productivity at renewal points allows firms to retain only those matches that exceed specific value thresholds, leading to non-random transitions between contract types and sectors.

Third, and most crucially for this setting, the model incorporates three distinct labor market segments: formal firms offering short-term contracts (STC), formal firms offering long-term contracts (LTC), and informal firms. Formal firms face institutional constraints such as firing costs $f$, which are binding only for long-term contracts. Short-term contracts cannot be terminated early, but are reassessed at the end of each contract period. Informal firms, by contrast, evade these regulatory costs and operate outside of formal protections, but they offer lower expected productivity and fewer job benefits.

The model captures contract dynamics through threshold decision rules: at the end of an STC period, a firm evaluates whether to terminate the match, renew the STC, or upgrade the worker to an LTC. These decisions are based on productivity thresholds $\tilde{z}_S$ and $\tilde{z}_L$, which respectively determine the lower and upper bounds of match continuation. Firms reassess matches at the conclusion of each STC period, whereas under LTC, firms may dismiss workers at any time conditional on incurring a severance cost. Thus, firms endogenously select contract type transitions and separations based on observed match quality and the timing structure of reassessment. A third threshold $\tilde{z}_{INF}$ governs separations in the informal sector. 

This extended structure enables the model to capture labor market segmentation, contract upgrading, job destruction, and firm-side screening—all of which are relevant for analyzing the employment effects of reforms that reduce firing costs and extend the use of short-term contracts. These components also allow for an internally consistent mapping between the reform and the patterns observed in the data, including shifts in formalization, contract upgrading, turnover rates, job tenure differentials, and wage dispersion across contract types.

\subsection{Time Framework and Agents}

We consider a continuous-time, infinite-horizon environment in which both firms and workers are infinitely lived and forward-looking. All agents discount future payoffs at a common discount rate $r > 0$, with present value defined using $\beta = \frac{1}{1 + r}$. Time is continuous and evolves indefinitely.

The economy features a unit measure of workers. Workers are ex ante homogeneous, but ex post heterogeneity arises through match-specific productivity draws. Unemployed workers search for jobs and aim to maximize the expected present discounted value of income flows from employment.

A mass $m$ of firms is active in the economy. Firms are heterogeneous in the types of contracts they offer and in their sectoral identity. In particular, firms may choose to operate in one of three environments: (i) the formal sector offering short-term contracts (STC), (ii) the formal sector offering long-term contracts (LTC), or (iii) the informal sector. Each firm seeks to maximize expected discounted profits, taking into account sector-specific constraints and costs.

Match-specific productivity $z$ is drawn from a common continuous distribution $F(z)$, and determines the output generated by a given firm-worker pair. Upon meeting a worker, a firm learns the productivity realization $z$, which then governs both the initial contract offer and future continuation decisions. Workers may be offered STC or LTC contracts in the formal sector, depending on the match quality, or they may work in the informal sector, which does not involve social security contributions or firing costs.

The model allows for endogenous transitions between employment states through contract reassessment and on-the-job search. Separation occurs when the expected value of continuation falls below an endogenous threshold that varies by sector and contract type.

\subsection{Sectoral Choice and Productivity}

Firms choose to operate in one of three environments: the formal sector with short-term contracts (STC), the formal sector with long-term contracts (LTC), or the informal sector. This decision is shaped by a trade-off between productivity advantages and compliance costs. Formal firms must adhere to labor market regulations, including requirements related to minimum wages, social security contributions, and restrictions on dismissal. In particular, formal firms offering long-term contracts must pay a firing cost $f > 0$ upon dismissal. In contrast, short-term contracts impose no firing cost but restrict termination during the contract period. Firms with STC arrangements reassess matches at the end of each contract cycle, while LTC firms can reevaluate their employment decisions continuously.

Informal firms evade these legal obligations entirely. They do not incur firing costs or pay social contributions, but they face structural disadvantages such as limited access to credit, lower capital investment, and reduced worker productivity. These productivity penalties are captured by a sector-specific parameter $a < 1$, which scales match output downward relative to the formal sector, where productivity is normalized to 1.

After meeting a worker, each firm draws a match-specific productivity $z$ from a continuous distribution $F(z)$, defined over a compact support. In the formal sector, total output from a match is $z$, while in the informal sector, it is $az$. Firms then decide whether to hire and under which contract type, with the goal of maximizing expected discounted profits. The firm’s subsequent decisions about continuation, promotion, or dismissal hinge on this productivity realization and the cost structures associated with each contract environment.

This setting captures the segmentation of labor markets observed in Benin and other developing economies, where formal and informal firms coexist and the use of different contract types reflects underlying productivity differences and regulatory constraints.

\subsection{Matching Process and Labor Market Frictions}

Labor market frictions arise from imperfect information and the time it takes for workers and firms to find suitable matches. The matching process follows a standard formulation: unemployed workers and vacancies are paired through a matching function $M(u, v)$, where $u$ is the measure of unemployed workers and $v$ is the number of posted vacancies. The function is increasing and concave in both arguments and satisfies constant returns to scale.

Labor market tightness is defined by the ratio $\theta = \frac{v}{u}$. Given this, the probability that a firm fills a vacancy is $q(\theta) = \frac{M(u,v)}{v}$, while the job-finding rate for an unemployed worker is $\theta q(\theta)$. These matching probabilities shape both vacancy posting behavior and the inflow into employment.

After matching, the firm observes the match-specific productivity $z$, which determines whether a job offer is extended and what contract is chosen. If hired under a short-term contract, the match proceeds for the fixed contract duration, after which the firm re-evaluates the match and decides whether to terminate, renew, or convert it to a long-term contract. This periodic reassessment allows firms to use STCs as screening mechanisms before committing to more rigid LTC arrangements.

In the case of long-term contracts, firms can dismiss workers at any time but must pay severance costs. Informal firms, in contrast, may sever employment freely, but their reduced productivity lowers the expected value of most matches. In all sectors, match continuation depends on whether the productivity draw exceeds a threshold that justifies the firm's expected surplus from retaining the worker.

Steady-state unemployment arises from the balance between the endogenous job destruction rate and the flow of new hires, both of which depend on these contract- and sector-specific productivity thresholds. As such, equilibrium unemployment reflects not only search frictions but also the institutional constraints and firm reassessment strategies embedded in the contract structure.

\subsection{Value Functions and Contract Environments}

The model operates in continuous time and separates each period into two conceptual stages: a search and matching sub-period, and a production sub-period. In the first stage, agents may form new matches or continue existing ones, with the possibility of on-the-job search. In the second, production occurs and match-specific decisions are made. Value functions for unemployment and for employment in informal jobs, short-term contracts (STC), and long-term contracts (LTC) characterize the recursive structure of the model.

\subsubsection*{Unemployed Workers}

Let \( \hat{U} \) denote the value of unemployment during the search sub-period. In this phase, the unemployed worker chooses which labor market \( j \in \{S, L, INF\} \) to search in, maximizing expected value:

\begin{equation}
\hat{U} = \max_{j \in \{S, L, INF\}} \left\{ 
p(\theta_j) x + (1 - p(\theta_j)) U 
\right\},
\end{equation}

where \( p(\theta_j) \) is the probability of matching in sector \( j \), and \( x \) is the value of being matched. If no match is formed, the worker continues in unemployment. In the production sub-period, the value of unemployment is:

\begin{equation}
U = b + \beta \mathbb{E}[\hat{U}'],
\end{equation}

where \( b \) represents the flow value of leisure or unemployment benefits, and \( \hat{U}' \) is the expected value of unemployment in the following search sub-period.

\subsubsection*{Informal Employment}

Let \( \hat{V}^{INF} \) denote the value of an ongoing informal match during the search sub-period. Informal workers search on the job at rate \( \lambda \). Their match is reassessed based on productivity, with separations occurring at threshold \( \tilde{z}^{INF} \). The value function is:

\begin{equation}
\hat{V}^{INF} = \max_x \left\{
\lambda p(\theta_{INF}(x))x + (1 - \lambda p(\theta_{INF}(x))) \left[ 
\int_{\tilde{z}^{INF}}^\infty V^{INF}(z') \, dF(z') + F(\tilde{z}^{INF}) U' 
\right]
\right\}.
\end{equation}

In the production sub-period, the match generates a flow of joint utility \( g(z) \), and its continuation value satisfies:

\begin{equation}
V^{INF}(z) = g(z) + \beta \mathbb{E}[\hat{V}'^{INF}],
\end{equation}

where \( \hat{V}'^{INF} \) is the expected value from the next search phase.

\subsubsection*{Short-Term Contracts (STC)}

Let \( \hat{V}^S \) denote the match value in the STC regime during the search sub-period. The worker may receive outside offers, and the firm either maintains or updates the match:

\begin{equation}
\hat{V}^S = \max_x \left\{
\lambda p(\theta_S(x))x + (1 - \lambda p(\theta_S(x))) \int_{-\infty}^\infty V^S(z') \, dF(z')
\right\}.
\end{equation}

During the production sub-period, the match yields current joint utility \( g(z) \), and at the end of the contract period, it is reassessed. The value function is:

\begin{equation}
V^S(z) = g(z) + \beta \left[
F(\tilde{z}^S) U' + \int_{\tilde{z}^S}^{\tilde{z}^L} V^S(z') \, dF(z') + \int_{\tilde{z}^L}^{\infty} V^L(z') \, dF(z')
\right].
\end{equation}

Here, matches are terminated if \( z < \tilde{z}^S \), renewed as STC if \( z \in [\tilde{z}^S, \tilde{z}^L) \), and upgraded to LTC if \( z \geq \tilde{z}^L \).

\subsubsection*{Long-Term Contracts (LTC)}

Let \( \hat{V}^L \) denote the value of a long-term match during the search sub-period. The worker may engage in on-the-job search, and the firm re-evaluates match continuation:

\begin{equation}
\hat{V}^L = \max_x \left\{
\lambda p(\theta_L(x))x + (1 - \lambda p(\theta_L(x))) \left[
\int_{\tilde{z}^L}^{\infty} V^L(z') \, dF(z') + F(\tilde{z}^L) U'
\right]
\right\}.
\end{equation}

In the production sub-period, the match generates joint utility \( g(z) \) and continues with expected value:

\begin{equation}
V^L(z) = g(z) + \beta \mathbb{E}[\hat{V}'^L].
\end{equation}

If productivity falls below the firing threshold \( \tilde{z}^L \), the match ends and the worker returns to unemployment, with the firm paying severance.

\subsection{Labor Market Equilibrium}

\subsubsection{Dismissal Decisions and Match Quality Thresholds}

Firms evaluate whether to enter the labor market and which type of contract or sector to operate in by comparing the expected value of profits to the cost of posting a vacancy. For each contract environment—short-term contract (STC), long-term contract (LTC), or informal (INF)—the firm solves a profit maximization problem conditional on a match with productivity \( z \).

For short-term contracts, matches cannot be terminated mid-contract. The firm earns joint utility \( g(z) \) during the production sub-period and expects to pay \( x \) as the negotiated wage or transfer that splits the surplus. The firm retains the match if it yields a positive continuation value. The expected profit from STC hiring is:

\begin{equation}
\Pi^S = \int_{\tilde{z}^S}^\infty \left( g(z) - x \right) dF(z),
\end{equation}

where \( \tilde{z}^S \) is the separation threshold such that matches with \( z < \tilde{z}^S \) are terminated after the STC period.

In long-term contracts, firms may terminate a match at any time but must pay a severance cost \( f \) upon dismissal. The match continues only if productivity is above the threshold \( \tilde{z}^L \). The expected profit under LTC is given by:

\begin{equation}
\Pi^L = \int_{\tilde{z}^L}^\infty \left( g(z) - x \right) dF(z) - f \cdot F(\tilde{z}^L),
\end{equation}

where the second term accounts for the expected cost of severance due to separations below \( \tilde{z}^L \).

For informal matches, which are not subject to firing costs or formal reassessment constraints, the profit function is similar to the STC case:

\begin{equation}
\Pi^{INF} = \int_{\tilde{z}^{INF}}^\infty \left( g(z) - x \right) dF(z).
\end{equation}

Entry into each sector is governed by a zero-profit condition, which equates the cost of posting a vacancy \( c \) to the expected return from filling the vacancy. This vacancy creation condition is:

\begin{equation}
c = q(\theta_j) \cdot \mathbb{E}[\Pi^j], \quad \text{for } j \in \{S, L, INF\},
\end{equation}

where \( q(\theta_j) \) is the probability of filling a vacancy in sector \( j \), and \( \mathbb{E}[\Pi^j] \) is the expected profit across the distribution of productivity draws conditional on hiring in that market.

Separation thresholds are determined by comparing the value of continuing the match to the worker’s outside option. For STC contracts, the threshold \( \tilde{z}^S \) is defined by the condition:

\begin{equation}
V^S(\tilde{z}^S) = U',
\end{equation}

indicating that the firm is indifferent between renewing the match and releasing the worker into unemployment. For long-term contracts, where separation involves severance, the threshold is lower, and satisfies:

\begin{equation}
V^L(\tilde{z}^L) = U' - f.
\end{equation}

For the informal sector, the separation rule is again based on the comparison to unemployment, without any firing cost. The productivity threshold satisfies:

\begin{equation}
V^{INF}(\tilde{z}^{INF}) = U'.
\end{equation}

These thresholds partition the productivity distribution into regions of dismissal and continuation, shaping job duration, turnover, and contract upgrading dynamics across sectors. The system of equations composed of the value functions, threshold conditions, and vacancy creation constraints jointly determine equilibrium sorting and labor market flows.

\subsection{Comparative Statics: Effects of Lowering Firing Costs}

This section analyzes the predicted general equilibrium effects of a reduction in firing costs \( f \) for long-term contracts (LTC), as enacted in Benin’s 2017 labor market reform. While stylized, the model provides useful insight into how such a policy shift can reconfigure contract usage, worker-firm matching, and wage-setting behavior in segmented labor markets.

The most immediate consequence of lowering \( f \) is an increase in the firm’s separation threshold for long-term contracts. The threshold \( \tilde{z}^L \) satisfies the indifference condition \( V^L(\tilde{z}^L) = U' - f \). As \( f \) falls, this right-hand side increases, meaning that firms require higher productivity realizations to retain a worker on an LTC. Firms become more willing to terminate matches that fall below this elevated threshold. This tightening of retention criteria introduces greater selection on match quality into the stock of long-term contracts.

The model also predicts an indirect increase in the separation threshold \( \tilde{z}^S \) for short-term contracts. Since firms face lower expected costs from upgrading STC workers to LTCs, they are more willing to convert high-productivity matches and less inclined to repeatedly renew STCs for borderline cases. The result is a narrowing of the productivity band for STC renewals. Matches that previously might have been retained as short-term are now either upgraded or terminated. This reinterpretation of STCs as screening devices—used to sort low- and high-productivity matches before committing to an LTC—leads to greater turnover among STC workers. This mechanism accounts for the post-reform decline in STC job tenure observed in the data.

The reform has no direct impact on the informal sector, as informal firms do not face firing costs. However, general equilibrium adjustments affect the composition of matches in the informal sector. As formal firms expand their hiring of higher-productivity workers—both because LTCs are less risky and because STCs now more effectively channel top matches toward LTCs—the pool of workers remaining in the informal sector becomes increasingly negatively selected. In this way, changes in formal sector policies indirectly affect the quality of informal employment, though this channel is not fully explored in the present model.

On the extensive margin, lowering \( f \) reduces the expected cost of job creation in the LTC segment. Since severance costs no longer significantly erode profits from uncertain matches, expected firm profits \( \mathbb{E}[\Pi^L] \) increase. Under the free-entry condition \( c = q(\theta_L) \cdot \mathbb{E}[\Pi^L] \), higher expected profits lead to more LTC vacancies and an increase in formal market tightness \( \theta_L \). With a higher vacancy-to-unemployment ratio, job seekers face better prospects in the LTC market, and formal sector employment expands.

This expansion in LTC vacancy posting also intensifies competition between formal firms to attract and retain high-productivity workers. In equilibrium, firms anticipate that the value of keeping top matches has increased—not only because of better expected productivity, but also because competitors are simultaneously expanding their hiring and upgrading thresholds. To secure and retain these workers, firms offer higher wages during the bargaining process. This endogenous wage pressure leads to observed wage increases, particularly among those in or near the promotion range from STC to LTC.

Taken together, the model predicts several reform-induced dynamics. Separation thresholds in both STC and LTC regimes rise, tenure falls among STC matches as screening becomes more selective, long-term job creation expands, formal market tightness increases, and wage gains are concentrated among high-productivity matches. These outcomes reflect re-optimization by firms facing lower dismissal costs and the resulting re-sorting of workers across contract types and sectors.

It is important to note that while the model captures key institutional features of segmented labor markets—such as contract duality and firing cost asymmetries—it abstracts from several realities. The model does not endogenize wage floors or enforcement heterogeneity, and assumes firms face a static productivity draw per match rather than learning over time. Nor does it incorporate equilibrium feedback from changes in household labor supply, job search behavior, or informal firm dynamics beyond their productivity penalty. As such, the comparative statics presented here should be interpreted as directional insights grounded in an internally consistent framework, rather than exact quantitative predictions.

Nonetheless, the mechanisms embedded in the model offer a coherent explanation for the patterns observed in the aftermath of Benin’s reform, including increased formalization, wage polarization, and contract upgrading. They also highlight the trade-offs policymakers face when loosening rigidities in the formal sector: gains in efficiency and formality may come alongside higher job turnover and more selective employment relationships.

\newpage
\section{Conclusion}\label{conclusion}

In this paper, I analyze the 2017 labor market reform in Benin, which aimed to increase the flexibility of the labor market by reducing firing costs and encouraging the use of short-term contracts. The effects of this reform were examined both theoretically and empirically, offering comprehensive insights into its impact on employment, contract types, job tenure, and wages.

From a theoretical perspective, the model I developed extends the Diamond-Mortensen-Pissarides (DMP) framework to better capture the specific labor market dynamics in Benin. The model incorporates endogenous job separation rates, which vary with economic conditions and worker-firm match quality, and distinguishes between the formal and informal sectors. In the formal sector, firms are subject to regulatory constraints like severance payments, whereas informal firms evade these regulations but operate at lower productivity levels. This theoretical framework provides a structure to understand how reduced firing costs affect labor market outcomes.

The model predicted that lowering firing costs would lead to an increase in formal sector vacancies by increasing the separation threshold. This, in turn, would increase formal sector employment but also reduce job tenure for lower-productivity workers. Empirical results corroborate these theoretical predictions. The reform resulted in a 2.6 percentage point increase in formal sector employment, equivalent to a 24.5\% increase compared to the pre-reform mean, and a 2.8 percentage point decrease in informal employment, translating to a 3.2\% decline. This reallocation suggests that lowering firing costs reduced barriers to entry in the formal sector, encouraging firms to expand formal job opportunities. However, overall employment remained relatively stable, with the probability of working at all decreasing slightly by 0.1 percentage points, or 0.10\%, indicating that the reform primarily shifted workers between sectors rather than significantly increasing total employment.

A major outcome of the reform was the substantial shift in contract types. The probability of having a permanent (long-term) contract in the formal sector increased by 23.2 percentage points, representing a 41.6\% increase for formal sector workers, and by 4.6 percentage points overall, an 88.4\% increase relative to the pre-reform mean. The model's Nash bargaining framework helps explain this result: firms, facing greater labor market flexibility, sought to retain high-productivity workers through long-term contracts to reduce turnover and maintain productivity. Empirical evidence supports this theoretical insight, as long-term contracts increased particularly among female workers (up 49.73\%), rural workers (up 59.55\%), and non-married individuals (up 57.61\%). These results reflect firms' strategies to secure valuable workers in a more competitive labor market.

However, the reform also led to a reduction in job tenure for short-term contract (STC) workers, as predicted by the theoretical model. The empirical results show that tenure for STC workers fell by 0.229 months, reflecting higher turnover due to the reduced firing costs. In contrast, tenure for workers on long-term contracts increased slightly by 0.153 months, as firms used these contracts to secure valuable employees. Overall, job tenure across all workers decreased by 0.158 months, highlighting the trade-off between increased employment flexibility and job stability.

Wage dynamics further confirm the predictions of the theoretical model. The reduction in firing costs allowed firms to offer higher wages, particularly for high-productivity workers on long-term contracts. Wages in the formal sector increased by 33.6 USD per month on average, representing a 28.34\% increase from the pre-reform average wage. Workers on long-term contracts saw an increase of 23.4 USD per month, or 13.86\%, while wages for short-term contract workers rose by 19.6 USD, reflecting a 35.48\% increase, though their wages remained lower overall. The Nash bargaining process in the model predicted that reduced firing costs would strengthen firms' ability to offer higher wages while securing productive workers.

In summary, the 2017 labor market reform in Benin successfully expanded formal sector employment, increased the prevalence of long-term contracts, and raised wages, particularly for high-productivity workers. However, the reform also introduced challenges related to job stability, especially for workers on short-term contracts, as evidenced by the reduction in tenure and higher turnover rates. 
It is important to recognize that some of the observed changes in labor market outcomes—particularly in average wages and contract durations—may reflect not only direct behavioral responses to the reform (i.e., true policy effects) but also shifts in the composition of workers entering the formal sector. While my empirical strategy accounts for several observable worker characteristics, I cannot fully separate these two channels using the available data. As such, the estimates presented should be interpreted as the combined result of policy-induced behavioral changes and compositional shifts in the workforce.

\clearpage
\printbibliography

\clearpage
\appendix
\section*{List of Tables}

\begin{table}[H]
    \centering
    \small
    \caption{Descriptive Characteristics of the Workforce in Benin and Other African Countries (Before and After Reform)}
    \label{descriptive_stat}
    \resizebox{.75\textwidth}{!}{
    \begin{tabular}{lcccccc}
        \toprule
        & \multicolumn{2}{c}{Comparison Group} & \multicolumn{2}{c}{Benin} & ttest \\
        \cmidrule(r){2-3} \cmidrule(l){4-5} \cmidrule(l){6-6}
        Survey Characteristics & Before & After & Before & After & P-value \\
        \midrule
        Working Age             &    34.94 &  35.69      &    34.47 &   35.15     &   0.067 \\
        Gender (Male)   &     0.54 &   0.53     &     0.47 &    0.48    &   0.000 \\
        Urban area      &     0.38 &   0.39     &     0.47 &  0.48      &   0.000 \\
        Household size  &     8.50 &    8.19    &     6.51 &    6.48    &   0.000 \\
        Married         &     0.67 &   0.66     &     0.72 &   0.67     &   0.000 \\
        No Education    &     0.63 &    0.62    &     0.60 &   0.60     &   0.041 \\
        Primary School & 0.19 &  0.17      &     0.19 &  0.19      & 0.951 \\
        Secondary School     &     0.16 &   0.19     &     0.18 &   0.18     &   0.000 \\
        Post Secondary School &     0.02 &   0.03     &     0.03 &     0.03   &   0.005 \\
        \midrule
        Observations & 110803&   113964&    16366&    17474&   127169\\
        \bottomrule
    \end{tabular}
    }
    \vspace{0.5cm} 
    \begin{minipage}{\textwidth}
        \footnotesize{
        \justifying
        \footnotetext{This table presents descriptive statistics for the sample of workers from Benin and a comparison group of other African countries, before and after the 2017 labor market reform. It compares key characteristics such as age, gender, education level, and economic status across the two groups. The p-values from t-tests are provided to assess whether the differences between the two groups are statistically significant. The sample size is shown at the bottom of the table, with 110,803 observations for the comparison group and 16,366 observations for Benin. Source: By Author.}
        }
    \end{minipage}
\end{table}

\begin{table}[H]
    \centering
    \small
    \caption{Employment Characteristics of Workers in Benin and Other African Countries (Before and After Reform)}
    \label{descriptive_stat_outcomes}
    \resizebox{.75\textwidth}{!}{
    \begin{tabular}{lcccc}
        \toprule
        & \multicolumn{2}{c}{Comparison Group} & \multicolumn{2}{c}{Benin} \\
        \cmidrule(r){2-3} \cmidrule(l){4-5}
         & Before & After & Before & After \\
        \midrule 
        \textbf{Share of Workers} &&&&\\
        \cmidrule(l){1-1}
        Total Employment &    97.53 &   97.35 &    98.40 &   98.35 \\
        Formal Sector Employment   &     12.71 &    12.41 &     10.59 &   13.34\\
        Informal Sector Employment &     84.82 &    84.94 &     87.81 &   85.01 \\
        \midrule
        \textbf{Contract Type} &&&&\\
        \cmidrule(l){1-1}
        Long Term Contract (Conditional)  &     74.36 &    72.57 &     55.75 &   81.87 \\
        Long Term Contract (Unconditional)  &     10.38 &    09.00 &     05.24 &   09.14 \\
        \midrule
        \textbf{Tenure in Last 12 Months} &&&&\\
        \cmidrule(l){1-1}
        Tenure Overall  &     8.220&   8.225 &     9.532 &  9.700\\
        Tenure Formal (Conditional)  &     9.434&   10.010 &     10.515 &  9.865\\
        Tenure Informal (Conditional)  &   8.271&    8.211 &   9.556 &  9.864\\
        \midrule
        \textbf{Under STC Workers} &&&&\\
        \cmidrule(l){1-1}
        Tenure Overall  &     8.050&   8.033 &     9.471 &  9.694\\
        Tenure Formal (Conditional)  &     8.725&   9.504 &     10.345 &  10.336\\
        Tenure Informal (Conditional)  &   8.271&    8.221 &   9.556 &  9.864\\
        \midrule
        Non-Employment Spell in Years &   1.407   & 1.460    &  1.361  &   1.552\\
        \midrule
        Monthly Earnings in Formal Sector (USD) &   148.663   &  155.321   &  118.602  &  93.491 \\
        \midrule
        Observations & 110803&   113964&    16366&    17474\\
        \bottomrule
    \end{tabular}
    }
    \vspace{0.5cm} 
    \begin{minipage}{\textwidth}
        \footnotesize{
        \justifying
        \footnotetext{
This table presents employment characteristics for workers in Benin and a comparison group of other African countries before and after the 2017 labor market reform. Columns for each group (comparison and Benin) show data from the periods before and after the reform. "Total Employment" reflects the percentage of individuals employed out of the total working-age population. "Formal Sector Employment" and "Informal Sector Employment" indicate the share of workers in each sector. "Long Term Contract (Conditional)" represents the percentage of workers with long-term contracts within the formal sector. "Long Term Contract (Unconditional)" provides the percentage of workers with long-term contracts across the entire sample, regardless of their employment sector. "Tenure in Last 12 Months" measures the average number of months workers have been in their current job over the last year. It is truncated above at 12 months. "Conditional" tenure metrics indicate the average tenure within specific sectors (formal or informal). Under STC Workers panel presents tenure statistics specifically for workers under short-term contracts. "Non-Employment Spell in Years" indicates the average duration of without employment among unemployed individuals, in years. "Monthly earnings in Formal Sector (USD)" reflects average earnings in the formal sector per month. By Author.}
}

    \end{minipage}
\end{table}

\begin{table}[H]
    \centering
    \caption{Two way fixed effect result: Probability of Working (Unconditional)}
    \label{working}
    \resizebox{.75\textwidth}{!}{
    \begin{tabular}{lccc}
        \hline
        & {(1) Working at all} & {(2) Formal sector} & {(3) Informal sector} \\
        \hline
        {Benin $\times$ 2019} &      -0.00147&              0.02629&             -0.02777\\
         se   &     (0.002)&     (0.004)&     (0.004)\\
          ci  &        [-0.006,0.003]&         [0.018,0.034]&       [-0.036,-0.019]\\
        \hline
        Observations & 258,599 & 258,599 & 258,599 \\
        Mean Pre-Treatment & 0.984 & 0.106 & 0.873 \\
        Percentage Effect & -0.10\% & 24.53\% & -3.21\% \\
        \hline
    \end{tabular}
    }
    \vspace{0.5cm} 
    \begin{minipage}{\textwidth}
        \footnotesize{
        \justifying
        \footnotetext{
This table presents the results of the impact of the 2017 labor market reform in Benin on the probability of working across different sectors. Column (1) shows the overall probability of working, column (2) focuses on the probability of working in the formal sector, and column (3) examines the probability of working in the informal sector. The coefficient for \textit{Benin $\times$ 2019} represents the estimated effect of the reform. Mean Pre-Treatment reflects the average probability in the treatment group before the reform. Standard errors are in parentheses, and 95\% confidence intervals are provided in brackets. All regressions are weighted using household weights and control for urban residence, age, gender, education, household size, marital status, industry type, and poverty level. Fixed effects for religion and commune are included. Robust standard errors are clustered at the household level.}
        }
    \end{minipage}
\end{table}

\begin{table}[H]
    \centering
    \caption{Probability of Having a Permanent Contract in the Formal Sector}
    \label{long_term_contract}
    \resizebox{0.8\textwidth}{!}{
    \begin{tabular}{lcc}
        \hline
        & {(1) Conditional Probability} & {(2) Unconditional Probability} \\
        \hline
       Benin $\times$ 2019&       0.232&                 0.046\\
       se     &     (0.020)         &     (0.004)         \\
        ci       &         [0.193,0.271]&         [0.039,0.053]\\
        \hline
        {Observations} & 32,758 & 258,599 \\
        {Mean Pre-Treatment} & 0.557 & 0.052 \\
        Percentage Effect & 41.65\% & 88.46\% \\
        \hline
    \end{tabular}
    }
    \vspace{0.5cm} 
    \begin{minipage}{\textwidth}
        \footnotesize{
        \justifying
        \footnotetext{
This table displays the impact of the 2017 labor market reform in Benin on the probability of obtaining a permanent contract. Column (1) presents the conditional probability, focusing solely on workers in the formal sector, while column (2) shows the unconditional probability, encompassing all workers irrespective of sector. The interaction term \textit{Benin $\times$ 2019} captures the estimated effect of the reform on the likelihood of holding a permanent contract. Mean Pre-Treatment reflects the average probability in the treatment group before the reform. Standard errors are in parentheses, and 95\% confidence intervals are provided in brackets. All regressions are weighted using household weights and include controls for urban residence, age, gender, education, household size, marital status, industry type, and poverty level. Fixed effects for religion and commune are also included. Robust standard errors are clustered at the household level.
}
        }
    \end{minipage}
\end{table}

\begin{table}[H]
    \centering
    \caption{Probability of Having a Permanent Contract by Heterogeneity Characteristics (Conditional on Working in the Formal Sector)}
    \label{long_term_contract_heterogeneity}
    \resizebox{\textwidth}{!}{
    \begin{tabular}{l*{8}{c}}
        \toprule
        & \multicolumn{2}{c}{Gender} & \multicolumn{2}{c}{Urbanicity} & \multicolumn{2}{c}{Age} & \multicolumn{2}{c}{Marital Status}\\
        & \multicolumn{1}{c}{(1) Male} & \multicolumn{1}{c}{(2) Female} & \multicolumn{1}{c}{(3) Urban} & \multicolumn{1}{c}{(4) Rural} & \multicolumn{1}{c}{(5) $\leq$ 35} & \multicolumn{1}{c}{(6) $>$ 35} & \multicolumn{1}{c}{(7) Married} & \multicolumn{1}{c}{(8) Not Married} \\        
        \midrule
        Benin $\times$ 2019 &  0.211&                 0.280&                 0.190&                 0.293&                 0.251&                 0.193&                 0.222&                 0.261\\
     se   & (0.023) & (0.038) & (0.025) & (0.038) & (0.026) & (0.027) & (0.024) & (0.042) \\
      ci      &         [0.165,0.256]&         [0.206,0.354]&         [0.142,0.239]&         [0.218,0.369]&         [0.201,0.301]&         [0.139,0.246]&         [0.175,0.268]&         [0.177,0.344]\\
      t-test   & \multicolumn{2}{c}{-1.48} & \multicolumn{2}{c}{-2.04} & \multicolumn{2}{c}{1.55} & \multicolumn{2}{c}{0.71} \\
      \hline
        Observations & 21,840 & 10,918 & 21,928 & 10,830 & 15,565 & 17,193 & 22,676 & 10,082 \\
        Mean Pre-Treatment & 0.554& 0.563& 0.585& 0.492& 0.519& 0.602& 0.588& 0.453\\
        Percentage Effect & 38.07\% & 49.73\%  & 32.48\% & 59.55\% & 48.36\% & 32.06\% & 37.76\% & 57.61\%\\
        \hline
    \end{tabular}
    }
    \vspace{0.5cm} 
    \begin{minipage}{\textwidth}
        \footnotesize{
        \justifying
        \footnotetext{
This table presents the estimated impact of the 2017 labor market reform in Benin on the probability of having a permanent contract, conditional on working in the formal sector, across various demographic and socioeconomic groups. Columns (1) and (2) represent the results for males and females, respectively; columns (3) and (4) show the results for urban and rural residents; columns (5) and (6) provide results by age group ($\leq$ 35 and $>$ 35); and columns (7) and (8) examine marital status (married and not married). The interaction term \textit{Benin $\times$ 2019} captures the estimated effect of the reform. The row labeled "t-test" presents the t-statistic ($t = \frac{\text{coefficient}_1 - \text{coefficient}_2}{\sqrt{\text{se}_1^2 + \text{se}_2^2}}$) for the difference in coefficients between each pair of subgroups (e.g., male vs. female). A t-test value closer to zero suggests no significant difference between the subgroups, while a larger absolute value indicates a more significant difference. Standard errors are in parentheses, and 95\% confidence intervals are provided in brackets. All regressions are weighted using household weights and include controls for urban residence, age, gender, education, household size, marital status, industry type, and poverty level. Fixed effects for religion and commune are also included. Robust standard errors are clustered at the household level.
}
        }
    \end{minipage}
\end{table}

\begin{table}[H]
    \centering
    \caption{Tenure (Last 12 Months) for all workers} 
    \label{tenure_12_months}
    \resizebox{\textwidth}{!}{
    \begin{tabular}{lccccccc}
        \hline
        &  & \multicolumn{2}{c}{Formal Sector (All)} & \multicolumn{2}{c}{Formal Sector (STC)} & \multicolumn{2}{c}{Informal Sector (STC)} \\
        \cmidrule(r){3-4} \cmidrule(r){5-6} \cmidrule(r){7-8}
        & (1) Overall Tenure & (2) Conditional& (3) Unconditional& (4) Conditional & (5) Unconditional & (6) Conditional & (7) Unconditional\\
        \hline
        Benin $\times$ 2019 &
         -0.158&      -0.118&       0.153&      -0.070&      -0.229&      -0.148&      -0.311\\
      se      &     (0.059)&     (0.108)&     (0.039)&     (0.231)&     (0.028)&     (0.060)&     (0.068)\\
     ci             &       [-0.273,-0.042]&        [-0.331,0.095]&         [0.075,0.230]&        [-0.524,0.384]&       [-0.285,-0.174]&       [-0.266,-0.031]&       [-0.445,-0.177]\\
   \hline
        Observations & 258,599&  32,758& 258,599&   8,536& 230,962& 219,677& 258,599 \\
        Mean Pre-Treatment & 9.533 & 10.515 & 0.990 & 10.345& 0.455& 9.556 & 8.543 \\
        \hline
    \end{tabular}
    }
    \vspace{0.5cm} 
\begin{minipage}{\textwidth}
    \footnotesize{
    \justifying
    \footnotetext{
This table presents the two way fixed effect results for worker tenure within the last 12 months, capturing how long workers have stayed in their current jobs at the time of the survey. Column (1) shows overall tenure across all workers and sectors. Columns (2) and (3) report tenure for all workers in the formal sector, where column (2) is conditional on being in the formal sector, and column (3) is unconditional, including all workers (with those not in the formal sector counted as having zero tenure). Columns (4) and (5) present tenure specifically for workers on short-term contracts (STC) in the formal sector, while columns (6) and (7) focus on tenure for short-term contract (STC) workers in the informal sector. The interaction term \textit{Benin $\times$ 2019} captures the estimated effect of the 2017 labor market reform in Benin on worker tenure. Standard errors are in parentheses, and 95\% confidence intervals are provided in brackets. Robust standard errors are clustered at the household level.}
    }
\end{minipage}
\end{table}

\begin{table}[H]
    \centering
    \caption{Non-Employment Spell in Years (How long have you been Unemployed)}
    \label{unemployment_spell}
    \resizebox{.65\textwidth}{!}{
    \begin{tabular}{lcc}
        \hline
        & (1) Conditional & (2) Unconditional\\
        & Unemployed Workers & All\\
        \hline
        Benin $\times$ 2019 & 0.397 & 0.010 \\
      se  & (0.228) & (0.005) \\
       ci     &        [-0.051,0.846]&        [-0.001,0.020]\\
        \hline
        Observations & 6,163 & 258,598 \\
        Mean Pre-Treatment & 1.360& 0.016\\
        \hline
    \end{tabular}
    }
    \vspace{0.5cm} 
    \begin{minipage}{\textwidth}
    \footnotesize{
    \justifying
    \footnotetext{
This table presents the two way fixed effect results for Non-Employment spell duration in years, measuring how long individuals have been unemployed. Column (1) shows the conditional effect on Non-Employment spell for unemployed individuals, meaning only workers who are currently unemployed are included. Column (2) displays the unconditional effect, where the entire sample is included, with employed individuals having a non-employment spell of 0. The interaction term \textit{Benin $\times$ 2019} captures the estimated impact of the 2017 labor market reform in Benin on the Non-Employment duration. Standard errors are in parentheses, and 95\% confidence intervals are provided in brackets. Robust standard errors are clustered at the household level.}
    }
\end{minipage}
\end{table}

\begin{table}[H]
    \centering
    \caption{Two way fixed effect Results for Monthly Earnings in the Formal Sector (in USD)}
    \label{wages_formal_sector_did}
    \resizebox{\textwidth}{!}{
    \begin{tabular}{lccccc}
        \toprule
        & \multicolumn{2}{c}{All Workers} & \multicolumn{2}{c}{Short Term Contract} & \multicolumn{1}{c}{Long Term Contract} \\
       \cmidrule(r){2-3} \cmidrule(r){4-5} \cmidrule(r){6-6}
        & (1) Conditional & (2) Unconditional & (3) Conditional & (4) Unconditional & (5) Unconditional \\
        & Formal Sector & All Workers & Formal Sector & All Workers & All Workers \\
        \midrule
        Benin $\times$ 2019 & 33.615 & 1.424 & 19.625 & -0.678 & 23.411 \\
        se   & (10.632) & (1.258) & (9.668) & (0.428) & (11.601) \\
        ci   & [12.720, 54.510] & [-1.046, 3.894] & [-0.607, 38.644] & [-1.526, 0.154] & [1.955, 50.320] \\
        \midrule
        Observations & 32,758 & 258,599 & 8,536 & 228,213 & 24,222 \\
        Mean Pre-Treatment & 118.602 & 11.168 & 55.236 & 2.461 & 168.911 \\
        \bottomrule
    \end{tabular}
    }
    \vspace{0.5cm} 
    \begin{minipage}{\textwidth}
        \footnotesize{
        \justifying
        \footnotetext{
This table presents the two way fixed effect results for monthly earnings in the formal sector, measured in USD after converting from FCFA using an exchange rate of 0.0017. Column (1) shows the conditional effect on wages for workers in the formal sector, while Column (2) presents the unconditional effect, considering all workers, including those not in the formal sector. Columns (3) and (4) present conditional and unconditional effects for workers with short-term contracts (STC) in the formal sector, while Column (5) shows the unconditional effect for workers with long-term contracts (LTC). The interaction term \textit{Benin $\times$ 2019} captures the estimated impact of the 2017 labor market reform on wages in the formal sector. Standard errors are in parentheses, and 95\% confidence intervals are provided in brackets. Robust standard errors are clustered at the household level.}
        }
    \end{minipage}
\end{table}

\section*{List of Figures}

\section*{Appendix Figures}

\begin{figure}[H]
    \centering
    \caption*{Appendix Figure A1: Informal employment per country}
    \label{informality_graph}    
    \includegraphics[width=\textwidth]{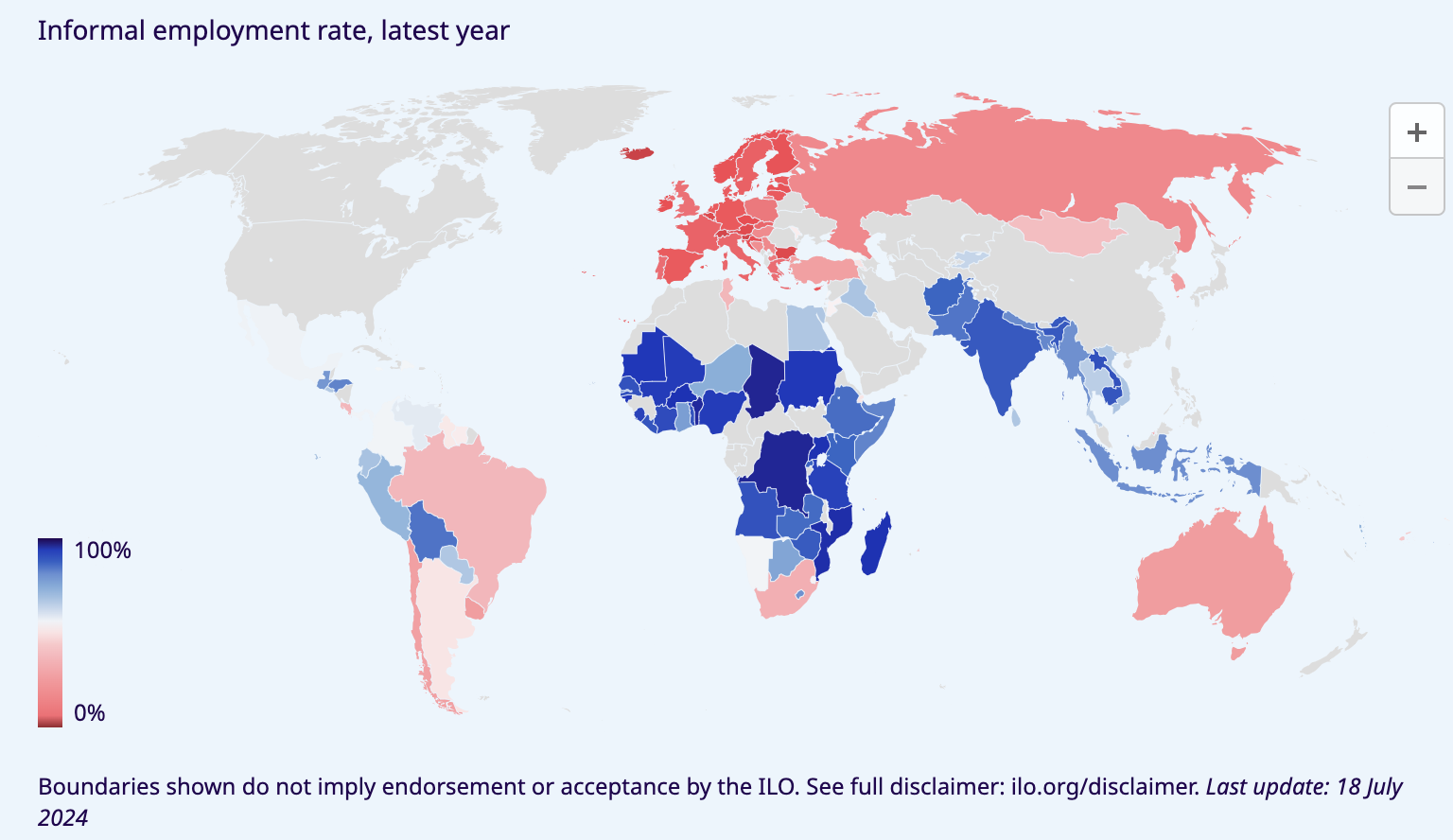}
    \captionsetup{justification=centering} 
    \vspace{0.5cm} 
    \begin{minipage}{\textwidth} 
        \footnotesize{
        \justifying
        \footnotetext{This map illustrates the latest global estimates of informal employment rates by country. Darker shades of blue indicate higher levels of informality, which tend to concentrate in Africa, Asia, and parts of Latin America. In contrast, many developed countries, particularly in Europe and North America, show lower levels of informal employment (in red and light blue shades). Benin is among the countries with one of the highest levels of informal employment, highlighting the relevance of labor market reforms aimed at formalization in the context of development. Data source: ILO, last updated July 18, 2024.}
        }
    \end{minipage}
\end{figure}

\begin{figure}[H]
    \centering
    \caption*{Appendix Figure A2: Distribution of Formal and Informal Firms in Benin}
    \label{sectors}
    \includegraphics[width=0.75\textwidth]{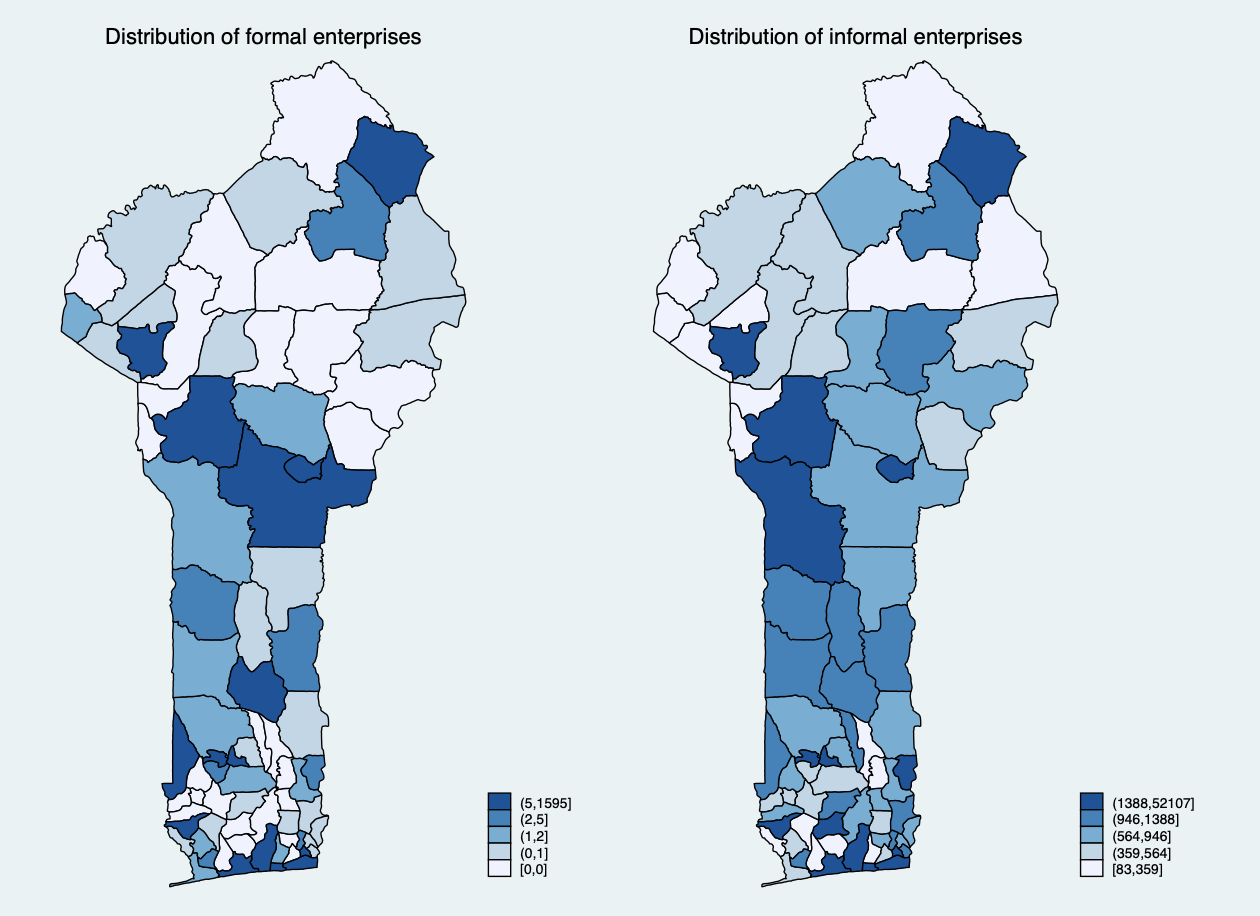}
    \vspace{0.5cm} 
    \begin{minipage}{\textwidth}
        \footnotesize{
        \justifying
        \footnotetext{The maps display the geographic distribution of formal (left) and informal (right) firms across Benin. Darker shades represent higher concentrations of firms, while lighter shades indicate lower concentrations. The formal sector is more prevalent in the southern regions, particularly around major cities, while the informal sector is distributed more evenly across the country. This visual contrast highlights the regional disparities in formal and informal economic activities. Source: By Author.}
        }
    \end{minipage}
\end{figure}

\begin{figure}[H]
    \centering
    \caption*{Appendix Figure A3: Workers Distribution in Formal versus Informal Employment by Industry}
    \label{industry_distribution}
    \includegraphics[width=\textwidth]{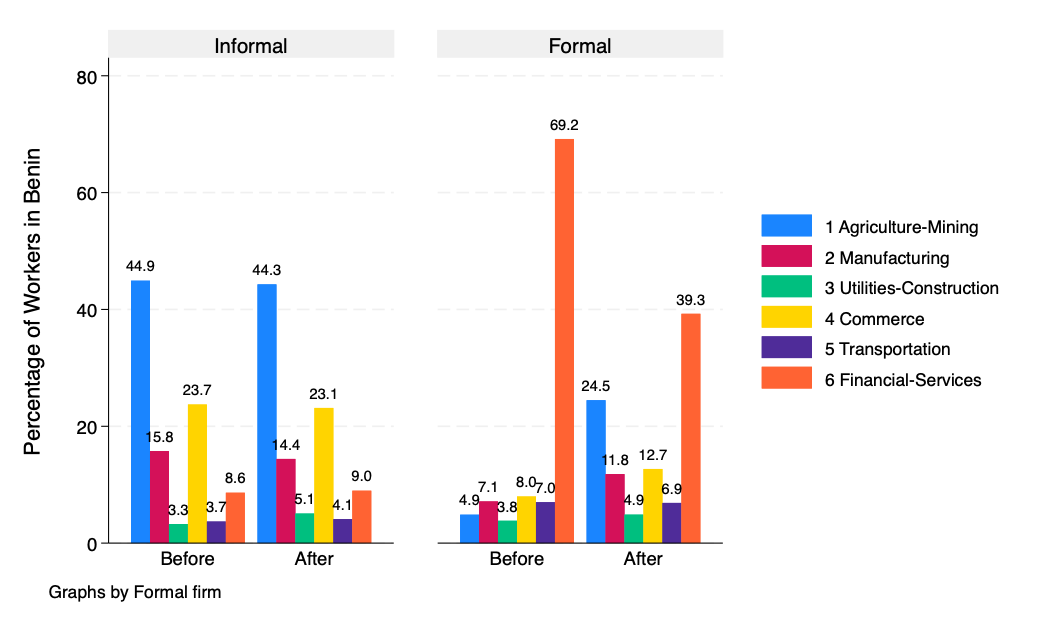}
    \vspace{0.5cm} 
    \begin{minipage}{\textwidth}
        \footnotesize{
        \justifying
        \footnotetext{This figure illustrates the distribution of workers across different economic sectors in Benin, comparing formal (right panel) and informal (left panel) employment before and after the labor market reform. The x-axis represents industries: Agriculture-Mining, Manufacturing, Utilities-Construction, Commerce, Transportation, and Financial Services. The y-axis shows the percentage of workers within each sector. Increases in formal employment are observed in all sectors except Financial Services, where formal employment decreased from 69.2\% to 39.3\% after the reform. Significant increases in formal employment are particularly notable in Agriculture-Mining (from 4.9\% to 24.5\%), Manufacturing (from 7.1\% to 11.8\%), Utilities-Construction (from 3.8\% to 4.9\%), and Commerce (from 8.0\% to 12.7\%). This visual highlights the impact of the reform on formalizing employment across most sectors. Source: By Author.}
        
        }
    \end{minipage}
\end{figure}

\begin{figure}[H]
    \centering
    \caption*{Appendix Figure A4: Number of workers in Formal versus Informal Employment by Industry}
    \label{industry_distribution_count}
    \includegraphics[width=\textwidth]{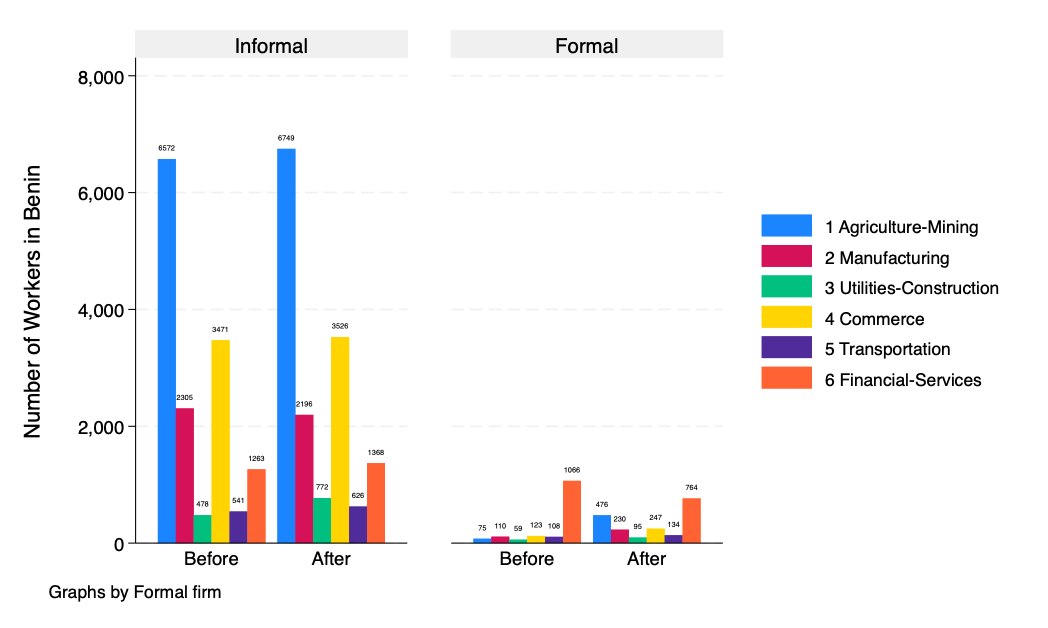}
    \vspace{0.5cm} 
    \begin{minipage}{\textwidth}
        \footnotesize{
        \justifying
        \footnotetext{This figure illustrates the number of workers across different economic sectors in Benin, comparing formal (right panel) and informal (left panel) employment before and after the labor market reform. The x-axis represents industries: Agriculture-Mining, Manufacturing, Utilities-Construction, Commerce, Transportation, and Financial Services. The y-axis shows the percentage of workers within each sector. This visual highlights the impact of the reform on formalizing employment across most sectors. Source: By Author.}    }
    \end{minipage}
\end{figure}

\begin{figure}[H]
    \centering
    \caption*{Appendix Figure A5: Workers Distribution Across Sectors in Benin}
    \label{sector_distribution_count}
    \includegraphics[width=\textwidth]{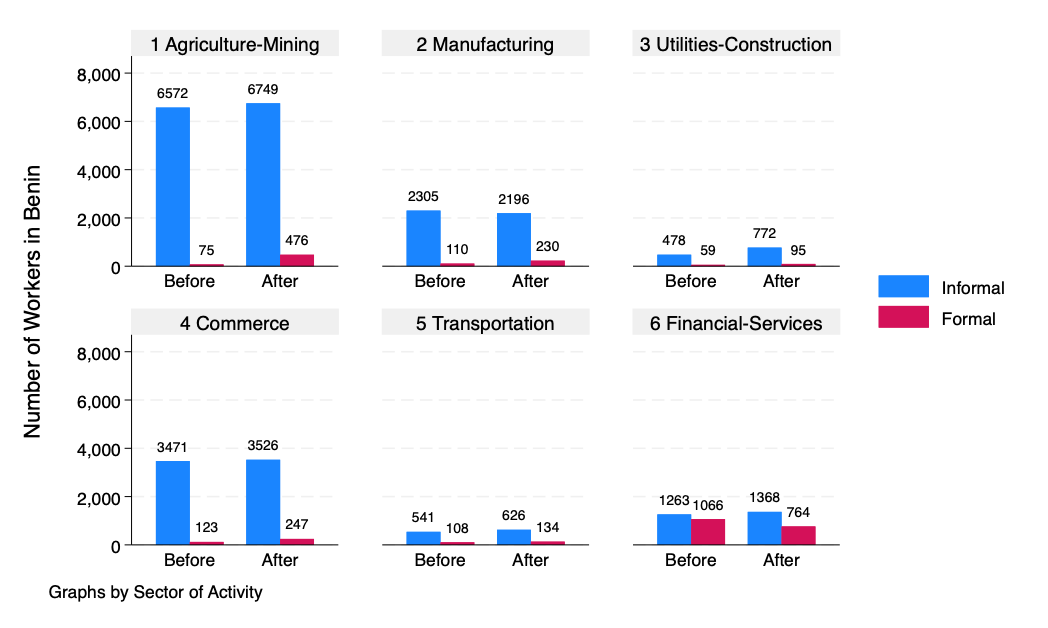}
    \vspace{0.5cm} 
    \begin{minipage}{\textwidth}
        \footnotesize{
        \justifying
        \footnotetext{This figure shows the number of workers across different economic sectors in Benin before and after the labor market reform. The x-axis displays two time periods labeled "Before" and "After" the reform. The y-axis represents the percentage of workers in each sector. Blue bars indicate informal employment, while red bars represent formal employment. Each subpanel corresponds to a specific industry: Agriculture-Mining, Manufacturing, Utilities-Construction, Commerce, Transportation, and Financial Services. This distribution provides insight into the sector-specific impacts of the reform on formal and informal employment in Benin. Source: By Author.}
        }
    \end{minipage}
\end{figure}

\begin{figure}[H]
    \centering
    \caption*{Appendix Figure A6: Event Study: Probability of Working}
    \label{working_all}
    \includegraphics[width=0.5\textwidth]{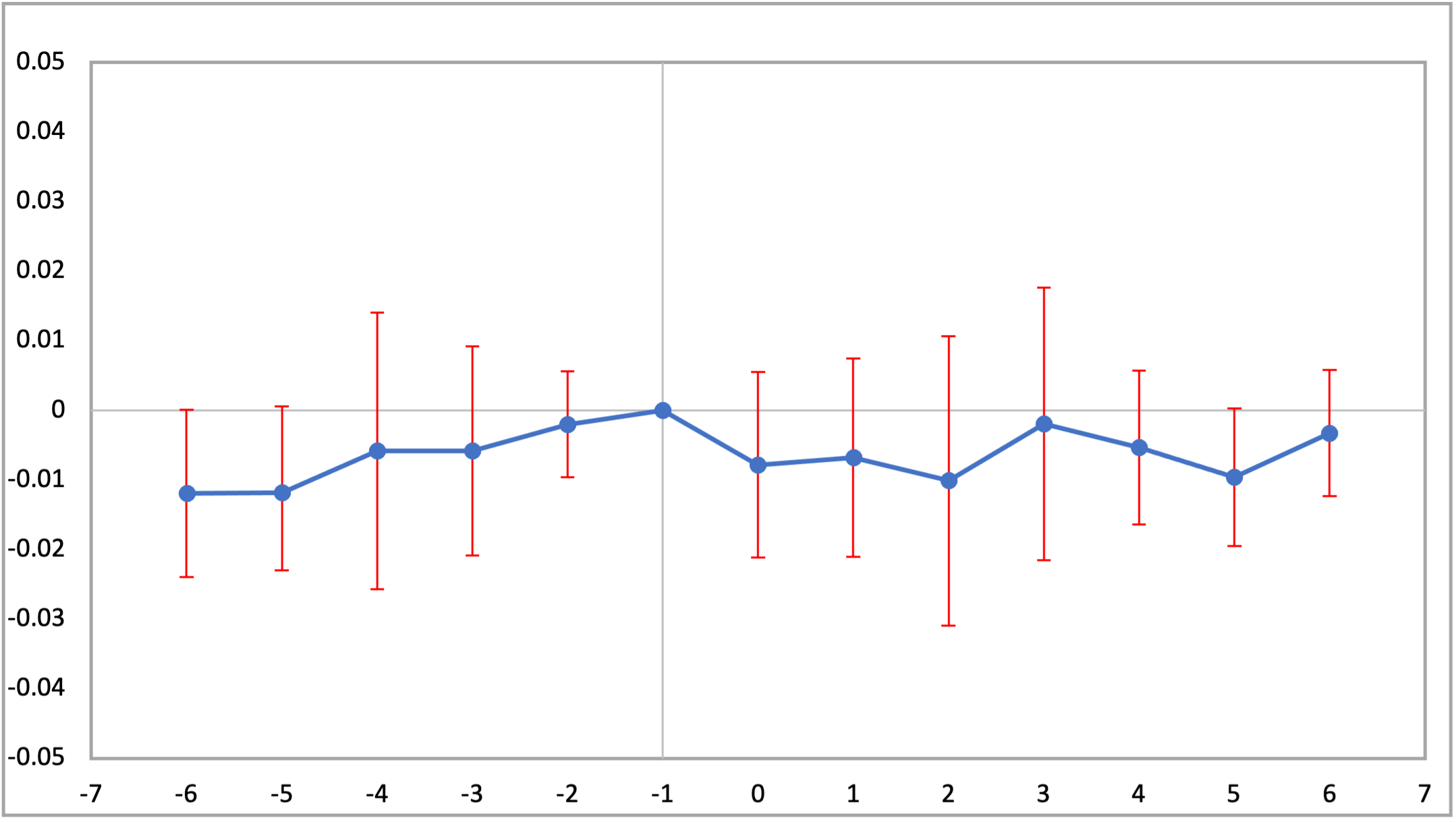}
    \vspace{0.5cm} 
    \begin{minipage}{\textwidth}
        \footnotesize{
        \justifying
        \footnotetext{This figure presents the event study analysis examining the probability of working before and after the labor market reform. The horizontal axis displays periods relative to the reform implementation: lead periods from –6 to –1 represent the months in which data was collected before the reform, with –1 as the reference period. Periods from 0 to 6 are lags, indicating the months of data collection following the reform. The y-axis shows the estimated change in the probability of working. The event study captures both short-term and long-term adjustments in employment. Although the LSMS data are cross-sectional, each wave was collected progressively over several months. The monthly event-time variable is constructed based on the interview date, with month –6 corresponding to January 2016 and month 0 to January 2019. Individuals are not tracked over time. The sample includes working-age individuals (15–64) who were either employed or unemployed and actively seeking work at the time of the survey.}
        }
    \end{minipage}
\end{figure}

\begin{figure}[H]
    \centering
    \caption*{Appendix Figure A7: Event Study: Probability of Working in the Formal Sector}
    \label{working_formal}
    \includegraphics[width=0.5\textwidth]{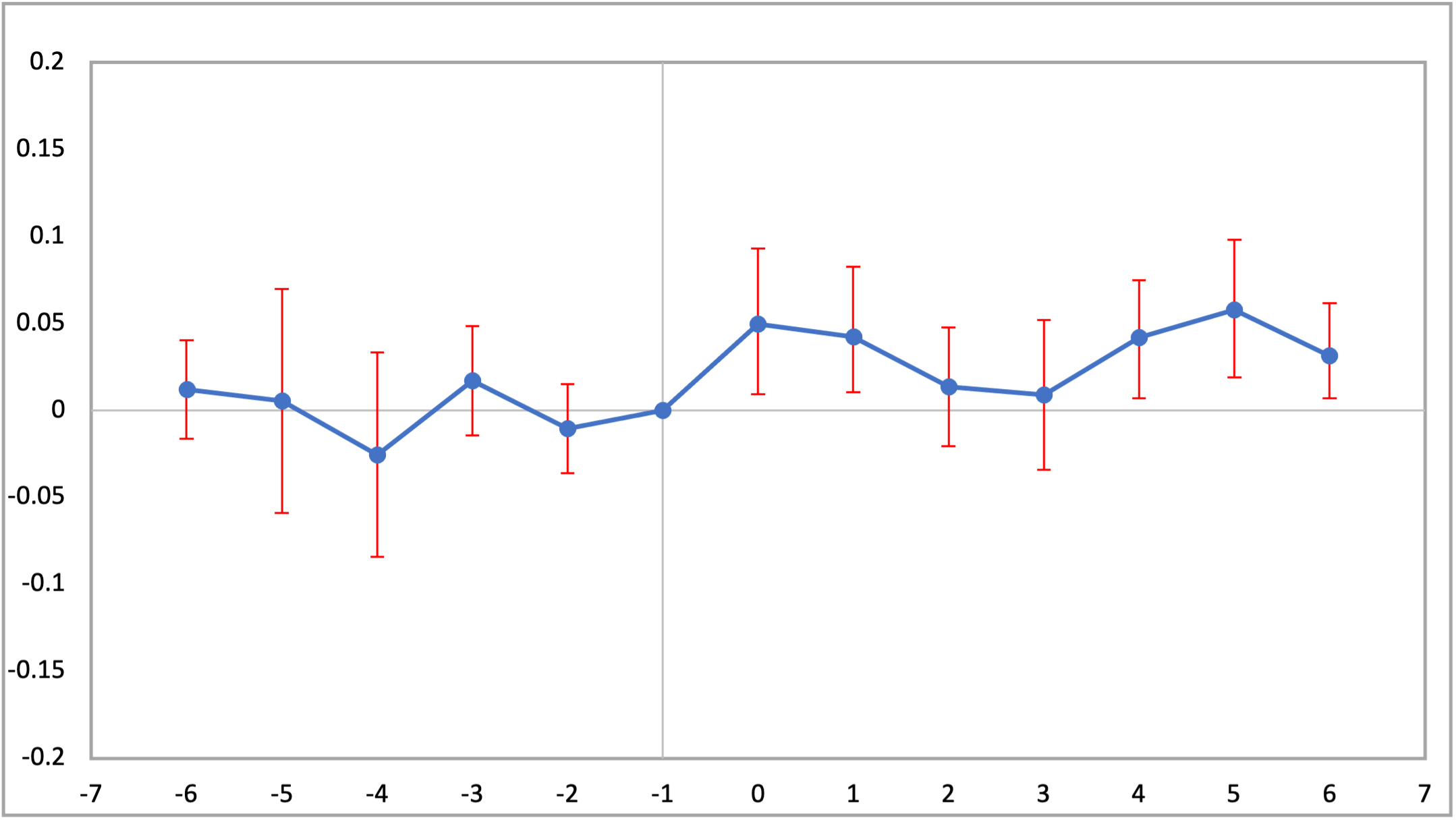}
    \vspace{0.5cm} 
    \begin{minipage}{\textwidth}
        \footnotesize{
        \justifying
        \footnotetext{This event study shows the estimated effect of the reform on the probability of working in the formal sector. The horizontal axis displays periods relative to the reform implementation: lead periods from –6 to –1 represent the months in which data was collected before the reform, with –1 as the reference period. Periods from 0 to 6 are lags, indicating the months of data collection following the reform. The y-axis indicates the variation in formal sector employment likelihood. Although the LSMS data are cross-sectional, each wave was collected progressively over several months. The monthly event-time variable is constructed based on the interview month, with month –6 corresponding to January 2016 and month 0 to January 2019. Individuals are not tracked over time. The sample includes working-age individuals (15–64) who were either employed or unemployed and actively seeking work at the time of the survey.}
        }
    \end{minipage}
\end{figure}

\begin{figure}[H]
    \centering
    \caption*{Appendix Figure A8: Event Study: Probability of Working in the Informal Sector}
    \label{working_informal}
    \includegraphics[width=0.5\textwidth]{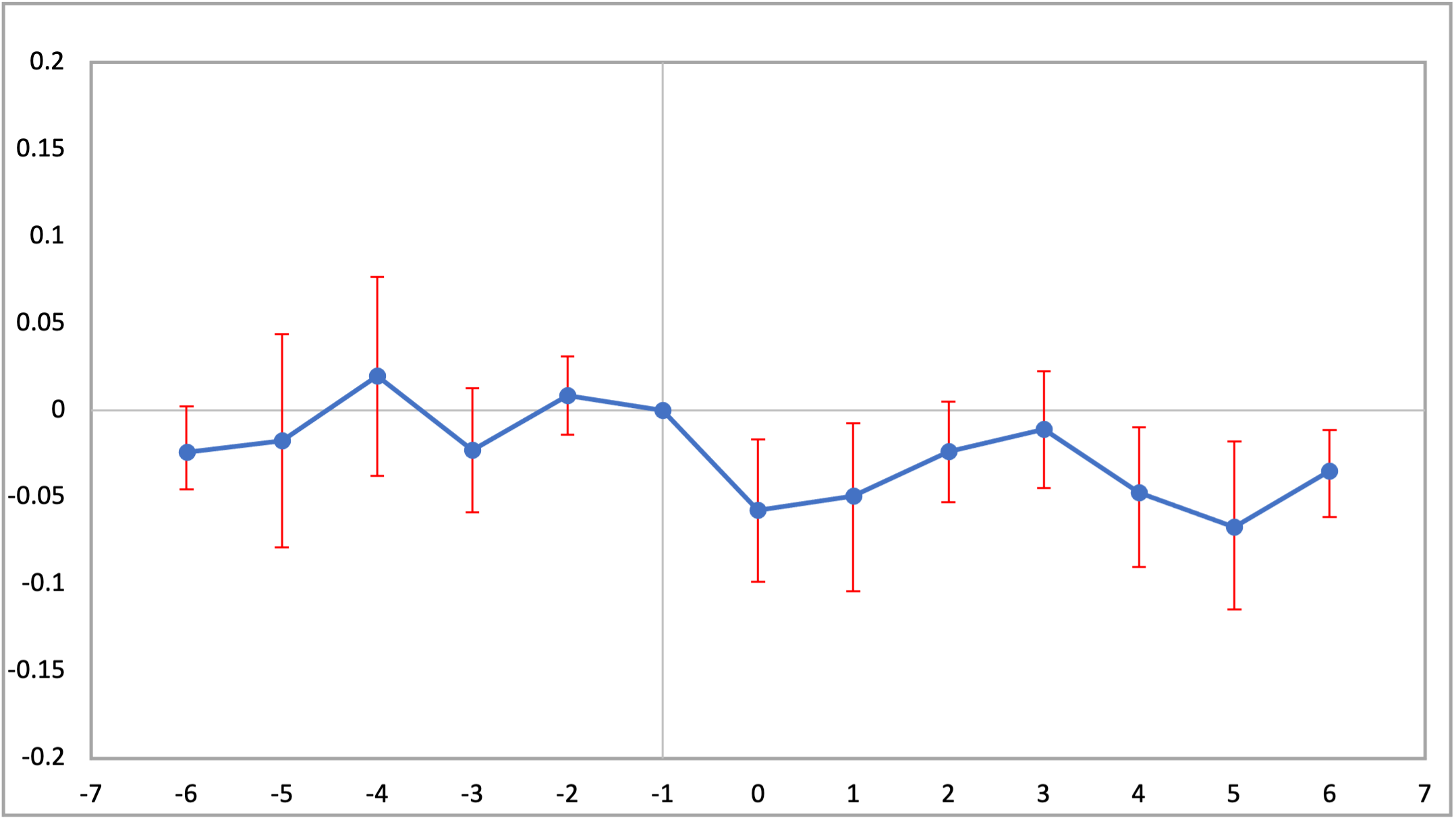}
    \vspace{0.5cm} 
    \begin{minipage}{\textwidth}
        \footnotesize{
        \justifying
        \footnotetext{This figure shows the impact of the reform on the probability of working in the informal sector. The horizontal axis displays periods relative to the reform implementation: lead periods from –6 to –1 represent the months in which data was collected before the reform, with –1 as the reference period. Periods from 0 to 6 are lags, indicating the months of data collection following the reform. The y-axis indicates changes in informal employment. The results suggest shifts in informal sector dynamics post-reform. Although the LSMS data are cross-sectional, each wave was collected progressively over several months. The monthly event-time variable is constructed based on the interview month, with month –6 corresponding to January 2016 and month 0 to January 2019. Individuals are not tracked over time. The sample includes working-age individuals (15–64) who were either employed or unemployed and actively seeking work at the time of the survey.}
        }
    \end{minipage}
\end{figure}

\begin{figure}[H]
    \centering
    \caption*{Appendix Figure A9: Event Study: Non-Employment Spell in Years (Unconditional)}
    \label{unem_spell_uncond_fig}
    \includegraphics[width=0.5\textwidth]{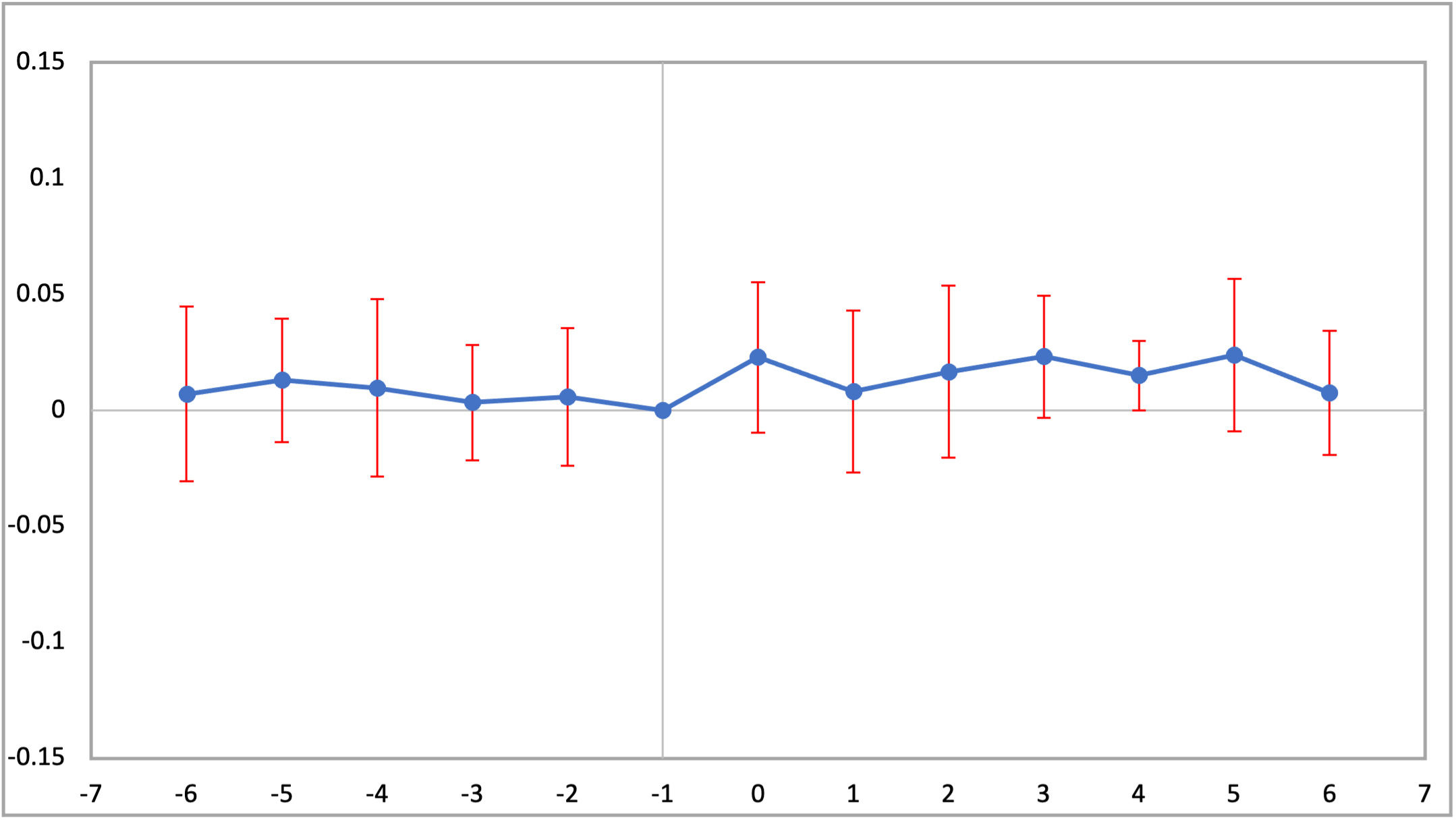}
    \vspace{0.5cm} 
    \begin{minipage}{\textwidth}
        \footnotesize{
        \justifying
        \footnotetext{
        This event study illustrates the average duration of unemployment spells in years across all workers, without restriction to currently unemployed individuals (unconditional sample). The x-axis represents time periods relative to the reform implementation, where –6 to –1 are lead periods, with –1 as the reference period. Periods from 0 to 6 are lags, representing months of data collection post-reform. The y-axis shows the deviation in unemployment spell duration from the reference period, allowing us to observe changes in unemployment duration trends following the reform. Although the LSMS data are cross-sectional, each wave was collected progressively over several months. The monthly event-time variable is constructed based on the interview month, with month –6 corresponding to January 2016 and month 0 to January 2019. Individuals are not tracked over time. The sample includes working-age individuals (15–64) who were either employed or unemployed and actively seeking work at the time of the survey.
        }
        }
    \end{minipage}
\end{figure}

\end{document}